\documentclass[11pt,a4paper]{article}
\usepackage{jcappub}
\usepackage[utf8]{inputenc}\usepackage[T1]{fontenc} 
\usepackage{amsmath}
\usepackage{amssymb}
\usepackage{graphicx}
\usepackage{slashed}
\usepackage{cases}
\usepackage[toc,page]{appendix}
\usepackage{empheq}
\usepackage[normalem]{ulem} 
\usepackage{xcolor} 
\usepackage{afterpage}
\usepackage{float}
\usepackage{bbold}
\usepackage{hypbmsec}
\usepackage{hyperref}
\usepackage{tikz}
\usepackage[compat=1.1.0]{tikz-feynman}  
\usepackage{slashed}                     
\usepackage{braket}
\usepackage{booktabs}
\usepackage{enumitem}
\usepackage{comment}

\usepackage{MnSymbol} 

\allowdisplaybreaks
\usepackage[export]{adjustbox}

\makeatletter
\newcommand\fake@math{}
\def\fake@math#1\){[math]}
\makeatother

\renewcommand{\d}{\mathrm{d}}

\newcommand{\rPS}{\bar{\rho}_{\mathrm{Ps}}}

\newcommand{\Neff}{N_\mathrm{eff}}

\begin{document}

\begin{flushright}
	{\large \tt MS-TP-24-33\\ CPPC-2024-09\\  IRMP-CP3-24-36}
\end{flushright}

\title{\boldmath  Towards a precision calculation of $\Neff$ in the Standard Model IV: Estimating the impact of positronium formation}

\author[a]{Tobias Binder,}
\emailAdd{tobias.binder@tum.de}
\author[b,a]{Marco Drewes,}
\emailAdd{marco.drewes@uclouvain.be}
\author[b]{Yannis Georis,}
\emailAdd{yannis.georis@uclouvain.be}
\author[c]{Michael~Klasen,}
\emailAdd{michael.klasen@uni-muenster.de}
\author[d]{Giovanni~Pierobon}
\emailAdd{g.pierobon@unsw.edu.au}
\author[d]{and Yvonne~Y.~Y.~Wong}
\emailAdd{yvonne.y.wong@unsw.edu.au}

\affiliation[a]{Physik Department, James-Franck-Stra\ss e~1,
Technische Universit\"at M\"unchen,\\
D--85748 Garching, Germany}
\affiliation[b]{Centre for Cosmology, Particle Physics and Phenomenology (CP3), 
Universit\'{e} catholique de Louvain, Chemin du Cyclotron 2, 
B-1348 Louvain-la-Neuve, Belgium}
\affiliation[c]{Institut f\"ur Theoretische Physik, Universit\"at M\"unster, Wilhelm-Klemm-Stra{\ss}e 9, D-48149 M\"unster, Germany}
\affiliation[d]{Sydney Consortium for Particle Physics and Cosmology, School of Physics, The University of New South Wales, Sydney NSW 2052, Australia}

\abstract{We present a first assessment of how the previously unexplored effect of positronium formation can impact on the value of the effective number of neutrino species in the Standard Model, $N_{\rm eff}^{\rm SM}$. 
Adopting a Yukawa form for the electrostatic potential, we discuss two possible scenarios that differ primarily in their assumptions about entropy evolution. The first, out-of-equilibrium scenario assumes that thermal corrections to the potential such as Debye screening prevent positronium from appearing until the temperature drops below a threshold. Once the threshold is reached, entropy generated in the QED sector from the equilibration process, if instantaneous, leads to a variation in $N_{\rm eff}^{\rm SM}$ of at most $|\Delta \Neff| \sim 10^{-4}$,
comparable to other uncertainties in the current benchmark value for~$N_{\rm eff}^{\rm SM}$. A more gradual formation could however yield a larger change. 
The second, equilibrium scenario assumes the QED sector to stay in equilibrium at all times.  In this case, we show that cancellations between the first, $s$-wave bound- and scattering-states contributions ensure that it is possible to evolve the system across the bound-state formation threshold without generating entropy in the QED sector. The corresponding change in $\Neff^{\rm SM}$ then closely matches the $\mathcal{O}(e^2)$ perturbative result derived in previous works and the $\mathcal{O}(e^4)$ contribution is capped at $|\Delta N_{\rm eff}| \lesssim 10^{-6}$.
We also comment on the impact of deviations from a pure Yukawa potential due to the presence of a thermal width.}

\maketitle

\section{Introduction}

The past few years have seen a renewed interest in the precision theoretical computation of the so-called effective number of neutrinos, $N_{\rm eff}$, in the context of the Standard Model (SM) of particle physics~\cite{Gariazzo:2019gyi,Akita:2020szl,Froustey:2020mcq,Bennett:2019ewm,Bennett:2020zkv,Cielo:2023bqp,Jackson:2023zkl,Drewes:2024wbw}.%
\footnote{In this work, the term ``Standard Model of particle physics'' is understood to include neutrino oscillations.}
Defined via an energy density ratio, 
\begin{equation}
N_{\rm eff}^{\rm SM}\equiv \frac{8}{7} \left(\frac{11}{4} \right)^{4/3} \left. \frac{\rho_\nu}{\rho_\gamma} \right|_{T/m_e \to 0, T/m_\nu \gg 0},
\label{eq:neffdefinition}
\end{equation}
the $N_{\rm eff}^{\rm SM}$ parameter quantifies the energy density contained in three ultra-relativistic SM neutrino species in the early universe after the electron-positron annihilation epoch ($T \sim 500$~keV) relative to the energy density in photons.  Changes to the value of $N_{\rm eff}$ impact directly on the universal expansion rate at early times and are in principle observable by a multitude of cosmological probes including the primordial light element abundances and the cosmic microwave background (CMB) anisotropies.  State-of-the-art calculations put the SM theoretical value at $N_{\rm eff}^{\rm SM} = 3.0440 \pm 0.0002$~\cite{Akita:2020szl,Froustey:2020mcq,Bennett:2020zkv},  although some dispute remains about the value of the last but one significant digit~\cite{Cielo:2023bqp,Jackson:2023zkl,Drewes:2024wbw}.
This number and the accompanying uncertainty are to be contrasted with the current observational constraint of $N_{\rm eff}^{\rm obs}=2.99 \pm 0.34$~(95\% C.I.) from the Planck CMB mission~\cite{Planck:2018vyg} and a potential future sensitivity of $\pm 0.02$ to $\pm0.03$ by CMB-S4~\cite{CMB-S4:2016ple}.

Within the SM, the physics that determines the precise value of $N_{\rm eff}$ takes place in the temperature range 
${\cal O}(10)~{\rm keV}\lesssim T \lesssim {\cal O}(10)~{\rm MeV}$, when the universe's energy is dominated by three families of SM neutrinos and a quantum electrodynamics (QED) sector comprising an electron/positron and photon bath.  The leading digit of $N_{\rm eff}^{\rm SM} \simeq 3$ accounts for three neutrino families. A basic description of neutrino decoupling from the QED plasma together with the QED equation of state to ${\cal O}(e^2)$ suffice to predict the leading-order correction of $\Delta N_{\rm eff} \simeq 0.04$~\cite{Dicus:1982bz,Dodelson:1992km,Heckler:1994tv,Hannestad:1995rs,Fornengo:1997wa,Lopez:1998vk,Esposito:2000hi,Mangano:2001iu,Birrell:2014uka,Grohs:2015tfy}.
Further refinements to the computation can be made to pin down the next digits.  These include finite-temperature corrections to the QED equation of state to ${\cal O}(e^3)$ and higher~\cite{Bennett:2019ewm}, distortions of the neutrino distribution from Fermi-Dirac statistics due to non-instantaneous neutrino decoupling and neutrino flavour oscillations~\cite{Dolgov:1997mb,Dolgov:1998sf,Mangano:2005cc,deSalas:2016ztq,Gariazzo:2019gyi,Akita:2020szl,Bennett:2020zkv,Froustey:2020mcq}, QED corrections to the weak interaction rates that govern neutrino decoupling~\cite{Cielo:2023bqp,Jackson:2023zkl,Drewes:2024wbw}, as well as the interplay between anisotropies and neutrino self-interactions~\cite{Hansen:2020vgm}.
See table~\ref{tab:Split} for a summary.   While these refinements are likely out of the reach of the forthcoming generation of cosmological probes, it is nonetheless useful to quantify their contributions to the theoretical value of $N_{\rm eff}^{\rm SM}$, so as to beat down the theoretical uncertainty in cosmological calculations and, in so doing, eliminate the need for error propagation in cosmological data analyses.

\begin{table}[t]
\centering
\begin{tabular}{lc}
\toprule
Standard-Model corrections to $N_{\rm eff}^{\rm SM}$ & Leading contribution \\
\midrule
$m_e/T_d$ correction& $+0.04$ \\
$\mathcal{O}(e^2)$ FTQED correction to the QED EoS& $+0.01$\\
Non-instantaneous decoupling+spectral distortion & $-0.006$\\
$\mathcal{O}(e^3)$ FTQED correction to the QED EoS& $-0.001$\\
Flavour oscillations & $+0.0005$\\
Type (a) FTQED corrections (thermal mass) to the weak rates  & $\lesssim 10^{-4}$\\
\midrule
Type (d) FTQED corrections (fermion loop) to the weak rates   & $\lesssim 10^{-5}$\\
 $\mathcal{O}(e^4)$ FTQED correction to the QED EoS & $3.5 \times 10^{-6}$ \\
 Electron/positron chemical decoupling & $\sim - 10^{-7}$ \\
\bottomrule
\toprule
Sources of uncertainty & \\
\midrule
Numerical solution by {\tt FortEPiaNO} & $\pm 0.0001$ \\
Input solar neutrino mixing angle $\theta_{12}$ &  $\pm 0.0001$ \\
\bottomrule
\end{tabular}
\caption{Leading-digit contributions from various SM corrections, in order of importance, thus far investigated that make up $N_{\rm eff}^{\rm SM}$. The first five corrections have been accounted for in precision computations leading to $N_{\rm eff}^{\rm SM}-3=0.0440 \pm 0.0002$~\cite{Akita:2020szl,Froustey:2020mcq,Bennett:2020zkv}; the rest are presently not part of the standard calculation as they fall below current numerical uncertainties.
Type~(a) FTQED corrections to the weak rates refer to thermal corrections to the electron mass which also change the electron phase space, while type~(d) corrections were considered in references~\cite{Jackson:2023zkl,Drewes:2024wbw} and correspond to a closed-fermion loop diagram that corrects, roughly speaking, the $\nu e$-scattering/annihilation matrix element. Electron/positron chemical decoupling refers to a small amount of entropy remaining in the electrons due to incomplete $e^+e^-$-annihilation after chemical decoupling at $T \simeq 20$~keV, which changes the equation of state of the QED sector~\cite{Thomas:2019ran}.
Table adapted from~\cite{Bennett:2020zkv,Drewes:2024wbw}; see also explanations therein.
}
\label{tab:Split}
\end{table}

In this work, we add to this list of refinements and consider for the first time the impact of positronium (Ps) formation on the theoretical value of $N_{\rm eff}^{\rm SM}$. Positronia, bound states of $e^+e^-$, may form when the electron/positron bath becomes sufficiently non-relativistic, i.e., at $T \lesssim m_e$. Although these bound states are unstable and subject to ionisation by the photon bath, should the conditions allow a {\it transient} population of such entities to exist and be produced and maintained in a state of {\it dynamical equilibrium} with the rest of the QED plasma at ${\cal O}(10)~{\rm keV} \lesssim T \lesssim m_e$, they could nonetheless alter the dynamics of the universe in ways that could affect~$N_{\rm eff}^{\rm SM}$.   Specifically, 
given that at these temperatures the bulk of the neutrino population has long since decoupled from the QED plasma, we expect a small population of positronium to alter $N_{\rm eff}^{\rm SM}$ predominantly via its impact on the QED equation of state, while corrections to the neutrino interaction rates are likely negligible.  Our main task in this work, therefore, is to assess  under what conditions positronium can form and be maintained in equilibrium in the time frame of interest, and its likely impact on the evolution of the QED plasma and hence $N_{\rm eff}^{\rm SM}$.

From a technical perspective, the greatest obstacle to estimating the impact of positronium is its large thermal width at the relevant temperatures, which renders its description as a collection of on-shell particles invalid.
This width arises from in-medium effects such as scattering and dynamically alters the properties of positronium states across the pertinent temperature range.   In practice, this means a na\"{i}ve Boltzmann-equation treatment---which relies on the on-shell assumption---is not applicable to our scenario, and tracking the system across the temperatures of interest necessitates a set of Kadanoff-Baym equations that also follow the four-point functions covering the out-of-equilibrium dynamics of off-shell positronium states in the medium. Deriving these equations and solving them numerically is well beyond the state of the art.  As a first estimate, we therefore consider instead two possible phenomenological models of positronium formation---out-of-equilibrium and adiabatic---in order to map out the relevant physics and appraise their likely impact on $N_{\rm eff}$ in a simplified way.

As a final note, we emphasise that our scenario of positronium formation differs from the common understanding of, e.g., hydrogen recombination or primordial nucleosynthesis.  In the latter two cases, the goal is to produce a frozen-out population of neutral hydrogen or nuclei free from ionisation or photodissociation.  On the other hand, our scenario does not require the positronium to avoid ionisation altogether as long as, once it is allowed to form, its formation rate is large enough to maintain a dynamical equilibrium with the free electrons/positrons in the plasma in the presence of ionising radiation. This situation is of course also possible for neutral hydrogen and nuclei at high temperatures, but is generally of negligible consequence to the evolution of the universe because of the highly-suppressed baryon number density during radiation domination.  The closest parallel we are aware of to our scenario is in fact the formation of positronium-like dark matter metastable bound states prior to dark matter freeze-out (e.g.,~\cite{vonHarling:2014kha,Vasilaki:2024fph}), although in that case the focus is not the equation of state of the universe, but the effect of bound states on freeze-out itself.

 The paper is organised as follows. We begin in section~\ref{sec:modellingneff} by asking how large an effect we can expect on $N_{\rm eff}^{\rm SM}$ if a population of positronium was produced out-of-equilibrium at some low temperature $T \lesssim m_e$, while section~\ref{sec:whendoesPsform} examines the conditions under which positronium can form and how fast they form. Section~\ref{sec:nonideal} considers the opposite limit in which positronium production takes place adiabatically and the QED plasma remains in equilibrium at all times.  We discuss future directions and conclude in section~\ref{sec:conclusions}.


\section{Out-of-equilibrium positronium formation and its impact on \texorpdfstring{$N_{\rm eff}^{\rm SM}$}{Neff}}
\label{sec:modellingneff}

\begin{table}[t]
\centering
\begin{tabular}{llc}
\toprule
Property & Expression at $T=0$ & Numerical value  \\
\midrule
Binding energy of energy level $n$ & $\Delta E_n = m_e \alpha^2/(4 n^2)$ & $6.8/n^2$~eV  \\
Mass & $m_{\rm Ps} = 2 m_e-\Delta E_1$ & $1.022$~MeV  \\
Bohr radius & $a_0 = 2/(m_e \alpha)$ & $536~{\rm MeV}^{-1}$ \\
Lifetime (for $n=1$): &&\\
$\quad \bullet$~Para-positronium ($s=0$) & $\tau_{0}=2/(m_e \alpha^5)$  & $0.1244$~ns \\
$\quad \bullet$~Ortho-positronium ($s=1$) & $\tau_1 = 9 \pi/[2(\pi^2-9) m_e \alpha^6] $ & $138.7$~ns \\
\bottomrule
\end{tabular}
\caption{Physical properties of the positronium at $T=0$ (see, e.g.,~\cite{Karshenboim:2005iy}).  The quantities $m_e\simeq 511$~keV and $\alpha = e^2/(4 \pi)\simeq 1/137$ are the electron mass and the fine structure constant, respectively. 
}
\label{tab:Psproperties}
\end{table}

Since positronium can only form when electrons and positrons are sufficiently non-relativistic, we focus our attention on the temperature epoch $10~{\rm keV} \lesssim T \lesssim m_e$.  Here, the universe's energy content comprises predominantly a QED plasma of photons and electrons/positrons held in thermal equilibrium at temperature $T$ by electromagnetic interactions, and three generations of SM neutrinos of temperature $T_\nu$, which have largely decoupled from the QED sector.  We assume as usual that the system is $CP$-symmetric with a vanishing chemical potential.    Table~\ref{tab:Psproperties} summarises some basic properties of the positronium at $T=0$ that will be useful for our discussion.

We estimate first in section~\ref{sec:entropy} the effect of instantaneous positronium production on $N_{\rm eff}^{\rm SM}$ via an entropy argument, before relaxing the instantaneous-production assumption in section~\ref{sec:relaxassumption}.


\subsection{Change in \texorpdfstring{$N_{\rm eff}^{\rm SM}$}{Neff} from instantaneous positronium production}
\label{sec:entropy}

Consider three epochs labelled by their scale factors, $a_1$, $a_2$, and $a_3$, corresponding respectively to 
\begin{enumerate}
\item the time of instantaneous neutrino decoupling, $T = T_\nu \simeq {\cal O}(1)$~MeV; 
\item positronium production, which we assume to be instantaneous and populate both $n=1$ singlet and triplet states, i.e., the para- and the ortho-positronium, at their equilibrium number densities at the same time; and 
\item some late time when all $e^+,e^-$, and Ps have annihilated to photons, where $N_{\rm eff}^{\rm SM}$ is defined via equation~\eqref{eq:neffdefinition}.  
\end{enumerate}
Except at $a_1$, the QED sector and neutrino temperatures differ, i.e., $T \neq T_\nu$.

Assuming that kinetic and chemical equilibrium holds at all times for all particle species {\it except} at the very instant positronium forms at $a_2$, entropy conservation tells us that
\begin{equation}
\begin{aligned}
s_1 a_1^3 & = s_{2-} a_2^3,\\
s_{2+} a_2^3 & = s_3 a_3^3,
\label{eq:entropy}
\end{aligned}
\end{equation}
where $s$ denotes the physical entropy density in the QED sector, the subscripts $2-$ and $2+$ mean just before and just after positronium formation, and we define $\delta s\equiv s_{2+}-s_{2-}$ to be the change in entropy at $a_2$.  Then, using the fact that the neutrino temperature scales as $T_\nu \propto a^{-1}$ after decoupling,
the relations~\eqref{eq:entropy} can be combined to give 
\begin{equation}
\frac{s_1}{T_{\nu,1}^3}\left(1+ \frac{\delta s}{s_{2-}} \right) =\frac{s_3}{T_{\nu,3}^3},
\end{equation}
and hence 
\begin{equation}
\frac{T_{\nu, 3}}{T_{3}} = \left[\frac{g_{*s,1}^{\rm QED}}{g_{*s,3}^{\rm QED}}  \left(1+ \frac{\delta s}{s_{2-}} \right)\right]^{-1/3} = \left(\frac{4}{11}\right)^{1/3} \left(1+ \frac{\delta s}{s_{2-}} \right)^{-1/3},
\end{equation}
where we have used $s \propto g_{*s}^{\rm QED} T^3$, and $g^{\rm QED}_{*s,1}=11/2$ and $g^{\rm QED}_{*s,3}=2$ denote, respectively the entropy degrees of freedom in the QED sector at neutrino decoupling and at $T/m_e \to 0$.%
\footnote{The number of entropy degrees of freedom in the QED sector at neutrino decoupling is in fact less than $g_{*s,1}^{\rm QED}=11/2$ because of a finite electron mass and finite-temperature corrections to the QED equation of state at $T \sim \mathcal{O}(1)$~MeV.  See reference~\cite{Bennett:2019ewm} for discussion.  This deviation is responsible for the leading-order correction to $N_{\rm eff}^{\rm SM}$, bringing its value from exactly 3 to 3.04.  However, as a perturbative, 1\%-level correction, its impact on the sub-leading correction due to positronium formation we wish to estimate in this work is negligible. Henceforth we shall always ignore this correction.  Functionally, this is the same as assuming that neutrinos decouple from the QED sector at $T/m_e \to \infty$.}
It follows straightforwardly that the neutrino and photon energy densities at late times are related via
\begin{equation}
\rho_{\nu,3} = 3 \times \frac{7}{8}  \left(\frac{4}{11}\right)^{4/3} \left(1+ \frac{\delta s}{s_{2-}} \right)^{-4/3} \rho_{\gamma,3}.
\end{equation}
Following the definition~\eqref{eq:neffdefinition} and writing $N_{\rm eff}^{\rm SM}=3+ \Delta N_{\rm eff}$, we then obtain
\begin{equation}
\Delta N_{\rm eff} =3 \left[\left(1+ \frac{\delta s}{s_{2-}} \right)^{-4/3}-1 \right] \simeq - 4 \frac{\delta s}{s_{2-}} 
\label{eq:dNeff}
\end{equation}
for the change in $N_{\rm eff}^{\rm SM}$ due to instantaneous positronium production.

It remains for us to estimate the change in entropy $\delta s$.  This can be achieved by matching the energy density in the QED plasma at the time just before positronium formation begins, with the energy density just after production is complete, i.e.,
\begin{equation}
\rho_e(a_2, T_2) + \rho_\gamma(a_2, T_2) = \rho_e(a_2, T_{2, {\rm new} }) + \rho_\gamma(a_2, T_{2,{\rm new}})  +\rho_{\rm Ps}(a_2, T_{2,{\rm new}}),
\label{eq:energymatch}
\end{equation}
and then solving for $T_{2, {\rm new}}$, where $T_{2, {\rm new}}$ is the new temperature the system settles to after an equilibrium positronium population appears instantaneously at $a_2$. Because positronium formation can only occur at low temperatures $T \ll m_e$, we expect $\rho_{\rm Ps} \ll \rho_\gamma$ and hence a small temperature change $\delta T \equiv T_{2,{\rm new}}-T_2$, $|\delta T/T_2| \ll 1$.  Then, equation~\eqref{eq:energymatch} can be expanded in $\delta T$ to give
\begin{equation}
\delta T  \simeq -\left. \frac{\rho_{\rm Ps}}{\frac{\partial}{\partial T}\left(\rho_e + \rho_\gamma +\rho_{\rm Ps}\right)}\right|_{T=T_2},
\end{equation}
and the change in entropy can be computed in the same limit from
\begin{equation}
\begin{aligned}
\delta s &\simeq s_{\rm Ps}(T_2) + \left. \frac{\partial}{\partial T}\left(s_e+s_\gamma + s_{\rm Ps} \right)\right|_{T=T_2} \delta T \\
&  = s_{\rm Ps}(T_2)- \left. \frac{\frac{\partial}{\partial T}\left(s_e+s_\gamma + s_{\rm Ps} \right)}{\frac{\partial}{\partial T}\left(\rho_e+ \rho_\gamma + \rho_{\rm Ps} \right)}\right|_{T=T_2} \rho_{\rm PS}(T_2).
\end{aligned}
\end{equation}
Noting that $s_\gamma \gg s_e,s_{\rm Ps}$  and $\rho_\gamma \gg \rho_e,\rho_{\rm Ps}$ at $T\ll m_e$, we use $s_\gamma \propto (4/3)\, T^3$ and $\rho_\gamma \propto T^4$ to further approximate
\begin{equation}
\delta s \simeq  s_{\rm Ps}(T_2)- \frac{\rho_{\rm Ps}(T_2)}{T_2} = \frac{P_{\rm Ps}(T_2)}{T_2},
\end{equation}
where $P_{\rm Ps}$ denotes the positronium pressure.  Substituting this $\delta s$ estimate into equation~\eqref{eq:dNeff} then gives a change in $N_{\rm eff}^{\rm SM}$ of
\begin{equation}
\Delta N_{\rm eff} \simeq -4 \times\frac{P_{\rm Ps}(T_2)/T_2}{s_\gamma(T_2) + s_e(T_2)} \simeq - \frac{45}{\pi^2} \frac{P_{\rm Ps}(T_2)}{T_2^4}<0,
\end{equation}
which we observe is always negative. This is consistent with the expectation that the main effect of positronium formation on the neutrino decoupling era is to slow down the redshifting of the QED bath and hence increase at late times the photon temperature relative to the neutrino one.

\begin{figure}
    \centering
    \includegraphics[width=0.75\linewidth]{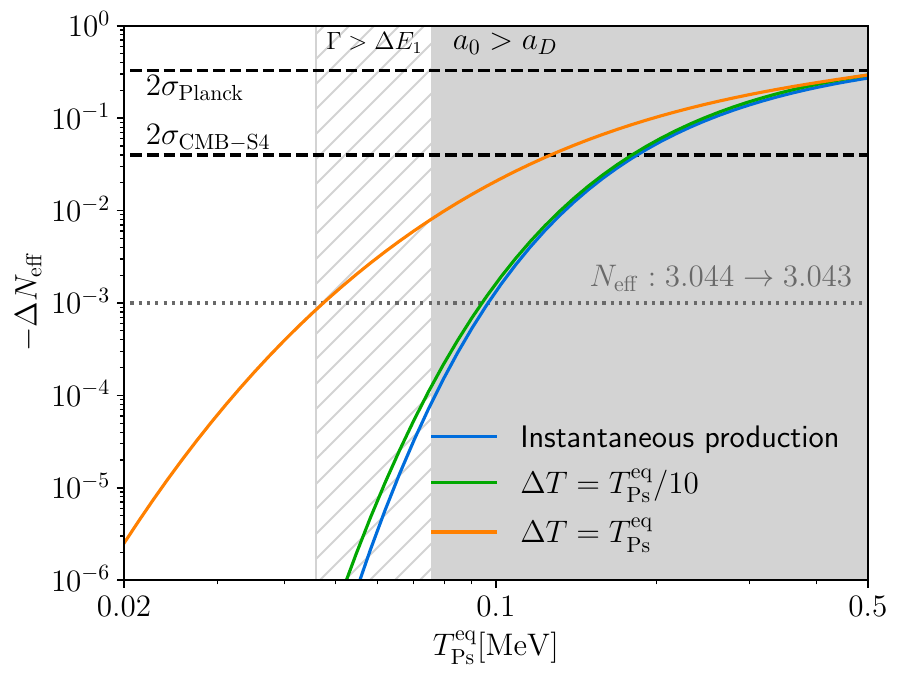}
    \caption{We display the (negative) change in $\Neff^{\rm SM}$, $-\Delta \Neff$, on a logarithmic scale as a function of the temperature $T_{\mathrm{Ps}}^{\mathrm{eq}}$ at which both $n=1$ positronium states reach equilibrium.  We use two different approximations to compute $\Delta \Neff$: The blue line represents estimates from the entropy argument of section~\ref{sec:entropy} assuming instantaneous positronium equilibration. The green and orange lines denote $\Delta \Neff$ from non-instantaneous production modelled with the parameterisation~\eqref{eq:parampositroniumformation} for the choices of $\Delta T/T^{\rm eq}_{\rm Ps} \in \{0.1,1\}$, respectively. Positronium formation is prevented by Debye screening in the grey-shaded region. In the hatched region, we expect scatterings with the plasma to prevent the formation of stable bound states, but more detailed computations are required to quantify the exact impact.  The three horizontal lines represent, from top to bottom, the current sensitivity of Planck at 95$\%$ C.L.~\cite{Planck:2018vyg}, the planned sensitivity of CMB-S4 \cite{CMB-S4:2016ple} at 95$\%$ C.L., and the last but one significant digit of the SM benchmark value $\Neff^{\rm SM}$ \cite{Gariazzo:2019gyi,Akita:2020szl,Froustey:2020mcq,Bennett:2019ewm,Bennett:2020zkv,Cielo:2023bqp,Jackson:2023zkl,Drewes:2024wbw}.} 
    \label{fig:tanh}
\end{figure}

Figure~\ref{fig:tanh} shows (blue line) $-\Delta N_{\rm eff}$ as a function of the instantaneous positronium production temperature $T_2$ (called $T_{\rm Ps}^{\rm eq}$ in the figure), assuming Bose-Einstein statistics for the positronium and an internal degree of freedom of $g_{\rm PS}=4$ (i.e., singlet+triplet).  Observe that in order to change $N_{\rm eff}^{\rm SM}$ by $|\Delta \Neff| \sim 10^{-3}$, instantaneous positronium equilibration needs to be achieved by $T \simeq 97$~keV.  This relatively high temperature (in comparison with the pertinent particle mass, $m_e$) is to be contrasted
with the impact of bound-state formation on freeze-out dark matter phenomenology~\cite{Feng:2009mn,vonHarling:2014kha,Vasilaki:2024fph}, where bound-state dynamics tends to enhance the annihilation cross-section, which delays freeze-out, thereby reducing the dark matter relic abundance. Translated to our system, an analogous cross-section enhancement would push electron/positron chemical decoupling---which takes place in the absence of bound-state dynamics at $T \simeq 20$~keV~\cite{Thomas:2019ran}---down to an even lower temperature.  The consequence is a more complete $e^+e^-$-annihilation prior to chemical decoupling and hence a smaller contribution from the residual electrons post decoupling to the entropy in the QED sector.  For $\Neff^{\rm SM}$, this means the correction due to a residual electron population would be {\it even smaller} in the presence of bound-state dynamics than the estimate of reference~\cite{Thomas:2019ran} -- $|\Delta \Neff| \sim 10^{-7}$ (see also table~\ref{tab:Split}), computed {\it without} bound-state considerations -- would suggest.%
\footnote{Reference~\cite{Thomas:2019ran} did not specify how they obtained the estimate $\Delta \Neff \sim - 10^{-7}$.  However, we find that we can reproduce this number using equation~\eqref{eq:dNeff} but with $\delta s/s$ reinterpreted to mean the fractional change in the QED entropy due to electrons/positrons immediately prior to chemical decoupling at $T \simeq 20$~keV.  Although the residual electrons remain kinetically coupled to photons after this time until a much lower temperature $T \lesssim 0.2$~eV, the miniscule baryon-to-photon number density ratio ($\sim 10^{-10}$) ensures that they do not alter the photon temperature evolution away from $T \propto a^{-1}$ in any appreciable way~\cite{Diacoumis:2018nbq}.  Thus, this estimate of $\Delta \Neff$ should remain to an extremely good approximation unchanged down to the CMB epoch.}

Finally, we emphasise again that, although the temperatures of interest far exceed the positronium binding energy $\Delta E_1$ of a few eV (see table~\ref{tab:Psproperties}), the presence of a minute positronium population is not precluded from {\it purely thermodynamic} considerations.  In fact, the positronium and free electron/positron  number densities, $n_{\rm Ps}$ and $n_{e^\pm}$, in our scenario are fully consistent with the Saha equation  (see, e.g., \cite{Binder:2021vfo}), $n_{e^-} /n_{\rm Ps} \propto (1/n_{e^+}) (m_e T/2 \pi)^{3/2}e^{-\Delta E_1/T}$ -- $\sim 2.1 \times 10^2$ at $ T \simeq 80$~keV and $\sim 9.7 \times 10^3$ at $T \simeq 50$~keV -- indicating that the vast majority of the electrons/positrons is, expectedly, unbound in the regime of interest. This also provides a justification for treating the electrons/positrons as a thermal bath.
Of course, at $T \gg \Delta E_1$ any individual positronium system is likely to be dissociated on times scales far shorter than the laboratory lifetimes given in table~\ref{tab:Psproperties}. However, for positronium to impact on the QED equation of state, it suffices that a transient population is maintained in dynamic equilibrium through efficient formation.
The questions of whether or not the formation rate is adequate to maintain this equilibrium, and, indeed, if positronia are even permitted to exist by finite-temperature requirements are deferred to section~\ref{sec:whendoesPsform}.

\subsection{Change in \texorpdfstring{$N_{\rm eff}^{\rm SM}$}{Neff} from non-instantaneous positronium production}
\label{sec:relaxassumption}

Relaxing the assumption of instantaneous equilibration should in general increase entropy production, thereby leading to a larger $|\Delta N_{\rm eff}|$. 
To account for this effect, we incorporate the contribution of positronium to the energy density of the universe directly in the continuity equation. As done in references~\cite{Bennett:2019ewm, Bennett:2020zkv, Drewes:2024wbw} and following the notation of~\cite{Mangano:2001iu, Mangano:2005cc}, the continuity equation for all particle species can be expressed as
\begin{align}
    \label{eq:continuity_adim}
     \frac{\d z}{\d x} = \frac{\left(\bar{\rho}_{\rm tot} - 3\bar{P}_{\rm tot}\right)/x - \partial \bar{\rho}_{\rm tot}/\partial x}{\partial \bar{\rho}_{\rm tot}/\partial z}
\end{align} 
in terms of the comoving variables $x= m_e\, a$,  $y=p\, a$, and $z=T\, a$, the rescaled total energy density $\bar{\rho}_{\rm tot} \equiv \bar{\rho}_e + \bar{\rho}_\gamma + \bar{\rho}_{\rm Ps} + \bar{\rho}_\nu =\rho_{\rm tot} a^4$, and the rescaled total pressure $\bar{P}_{\rm tot} = P_{\rm tot} a^4$.

In presence of photons, electrons and neutrinos only, the continuity equation~\eqref{eq:continuity} takes the standard form \cite{Mangano:2001iu, Drewes:2024wbw}
\begin{align}
\label{eq:continuity}
\frac{\d z}{\d x} &= \frac{\tau J_{+}(\tau)- 1/(2z^3)\d \bar{\rho}_\nu/\d x +G_1(\tau)}{\tau^2 J_{+}(\tau)+Y_{+}(\tau)+2\pi^2/15+G_2(\tau)},
\end{align} 
where $\tau=x/z=m_e/T$, the functions $J_{+}$ and $Y_{+}$ express the evolution of the electron/positron distribution across the annihilation epoch in the ideal gas limit, while 
$G_1$ and $G_2$ incorporate corrections to the QED equation of state due to finite temperature effects.  Explicit expressions for $J_{+}, Y_{+}, G_1$ and $G_2$ can be found in appendix~\ref{sec:continuitysupplement}. 
The quantity ${\rm d} \bar{\rho}_\nu/{\rm d} x$ corresponds to energy exchange between the neutrino and the QED sectors and can be set to zero if, as in section~\ref{sec:entropy}, we approximate neutrino decoupling to take place at $T/m_e \to \infty$.  Then, under this approximation, our variable $z$ is equivalently the photon-to-neutrino temperature ratio, i.e., $z=T/T_{\nu}$,%
\footnote{This temperature ratio interpretation of the variable $z$ is strictly true only when the neutrinos have completely decoupled from the QED sector.}
and evolving equation~\eqref{eq:continuity} from $x \to 0$ to $x \to \infty$ for an initial condition $z(x\to 0)=1$ allows us to estimate $\Neff^{\rm SM}$ from the relation 
\begin{equation}
    N_{\rm eff}^{\rm SM}- 3=3\left[\left(\frac{z_{\rm ref}}{z_{\rm fin}}\right)^4-1\right],
\end{equation} 
where we define $z_{\rm ref} \equiv (11/4)^{1/3}$. Note that if there is non-zero energy transfer between the neutrino and the QED sectors, i.e., ${\rm d} \bar{\rho}_\nu/{\rm d} x \neq 0$, then $z_{\rm fin}$ alone does not capture the change in $\Neff^{\rm SM}$---we also need to include contributions from changes in $\bar{\rho}_\nu$.

The continuity equation \eqref{eq:continuity_adim} can be straightforwardly generalised to include contributions from a positronium population of energy density and pressure $\bar{\rho}_{\rm Ps}$ and $\bar{P}_{\rm Ps}$,  which will generally need to be computed from a Boltzmann equation.
For our purpose of establishing an estimate, however,
we may model the evolution of $\bar{\rho}_{\rm Ps}$ as 
\begin{equation}
\label{eq:posformation_modelling}
    \bar{\rho}_{\rm Ps}(x,y,z)=\sigma(\tau)\bar{\rho}_{\rm Ps}^{\rm eq}(x,y,z)=\sigma(\tau)  \frac{g_{\rm Ps}}{2\pi^2}\int {\rm d}y\, y^2\; \frac{\sqrt{y^2+x_{\rm Ps}^2}}{\exp(\sqrt{y^2+x_{\rm Ps}^2}/z)-1} ,
\end{equation} 
with $x_{\rm Ps} = (m_{\rm Ps}/m_e)x \simeq 2x$, and the positronium population is assumed to have a thermal distribution of the same temperature $T$ as the QED plasma but rescaled by a smooth transfer function $\sigma(\tau) = f_{\rm Ps}/f_{\rm Ps}^{\rm eq}= n_{\rm Ps}/n_{\rm Ps}^{\mathrm{eq}}$ that depends only on $T$. The same ansatz for the positronium distribution function is assumed to model the evolution of the pressure $\bar{P}_{\rm Ps}$. Then, 
we find 
\begin{equation}
 \frac{1}{2z^3} \left(\frac{\rPS-3\bar{P}_{\mathrm{Ps}}}{x}  - \frac{\partial \bar{\rho}_{\mathrm{Ps}}}{\partial x}\right) 
=
\frac{g_{\rm Ps}}{4}  \left[\tau \left(\frac{m_{\rm Ps}}{m_e}\right)^2   J_{-}\left(\frac{m_{\rm Ps}}{m_e}\tau\right)- z W_-\left(\frac{m_{\rm Ps}}{m_e}\tau\right) \partial_x \right] \sigma(\tau),
\end{equation}
and 
\begin{equation}
\begin{aligned}
 \frac{1}{2z^3}\frac{\partial \rPS}{\partial z} =  & \; \frac{g_{\rm Ps}}{4} \left[ \tau^2\left(\frac{m_{\rm Ps}} 
   {m_e}\right)^2 J_{-}\left(\frac{m_{\rm Ps}}{m_e}\tau\right)+\frac{m_{\rm Ps}}{m_e} Y_{-}\left(\frac{m_{\rm Ps}}{m_e}\tau\right) 
    + zW_-\left(\frac{m_{\rm Ps}}{m_e}\tau\right) \partial_z \right] \sigma(\tau), 
  \end{aligned}
\end{equation}
where $J_{-}$ and $Y_{-}$ are the Bose-Einstein equivalent of $J_{+}$ and $Y_{+}$, and $W_-(\tau)$ is a new function, all of which can be found in appendix~\ref{sec:continuitysupplement}.  Setting $m_\mathrm{Ps} \simeq 2m_e$ and $g_\mathrm{Ps}=4$ and using the property $\partial_z \sigma = -(x/z) \partial_x \sigma$, we obtain for the continuity equation
\begin{align}
\label{eq:continuityeq_ps_sigma}
\frac{\d z}{\d x} &= \frac{\tau[J_{+}(\tau)+4\sigma(\tau)J_-(2\tau)]-z\partial_x\sigma(\tau)W_-(2\tau)}{\tau^2\left[J_{+}(\tau)+4\sigma(\tau)J_-(2\tau)\right]+Y_{+}(\tau)+Y_-(2\tau)-x\partial_x\sigma(\tau)W_-(2\tau)+2\pi^2/15},
\end{align} 
having dropped for simplicity the terms proportional to $G_{1,2}$ and to the neutrino abundance.%
\footnote{We have checked explicitly that adopting a more realistic modelling that includes $G_{1,2}$ and ${\rm d}\bar{\rho}_\nu/{\rm d}x$ has no substantial impact on our estimate of $\Delta N_{\rm eff}$ from a transient positronium population.}

For the transfer function $\sigma(\tau)$, we generally expect the energy density of a species far from equilibrium to scale as $\rho \propto 1-e^{-\int \Gamma_{\rm prod}/(Hx) \d x}$,
where $\Gamma_{\rm prod}$ is the production rate and $H$ the Hubble expansion rate.  We therefore adopt as a first guess the {\it ansatz}
\begin{align}
\label{eq:parampositroniumformation}
    \sigma(T) = \frac{1}{2}\left[1+\tanh\left(\frac{T^{\rm eq}_{\rm Ps}-T}{\Delta T}\right)\right],
\end{align} 
where the width of the $\tanh$ function $\Delta T$ parameterises the effective positronium equilibration timescale, and $T=m_e/\tau = m_e (z/x$) in terms of the dimensionless variables. 
We show in figure~\ref{fig:tanh} (green and orange lines) the expected $\Delta \Neff$ for two choices of $\Delta T\in\{T^{\rm eq}_{\rm Ps}/10,T^{\rm eq}_{\rm Ps}\}$, which serve as benchmarks for a fast and a slow equilibration  respectively. As expected, the faster the equilibration, the more closely the change in $\Neff^{\rm SM}$ approaches the estimates from entropy arguments (section~\ref{sec:entropy}); in fact, taking the limit $\sigma(\tau)\rightarrow \Theta\left[(T^{\rm eq}_{\rm Ps}-m_e/\tau)/\Delta T\right]$, where $\Theta$ is a Heaviside function, and adjusting the temperature to account for the sudden increase in entropy at $T = T^{\rm eq}_{\rm Ps}$ allow us to recover the estimates of section~\ref{sec:entropy} exactly.   We have also tested other functional forms of $\sigma$, and find that as long as the model contains an exponential dependence on $(T^{\rm eq}_{\rm Ps}-T)/\Delta T$, it will yield a $\Delta \Neff$ similar to the model~\eqref{eq:parampositroniumformation}.


\section{When does positronium form?}
\label{sec:whendoesPsform}

We have seen in figure~\ref{fig:tanh} that the change in $\Neff^{\rm SM}$ due to out-of-equilibrium positronium formation is a rapidly changing number with respect to the positronium equilibration temperature $T_{\rm Ps}^{\rm eq}$. We now need to estimate $T_{\rm Ps}^{\rm eq}$ and how fast positronium equilibration can be achieved in the presence of the QED plasma. 

Generally speaking, the spectrum and properties of bound states are encoded in the dressed spectral functions in the medium, which evolve dynamically (along with the occupation numbers) and are heavily affected by the surrounding plasma. 
To rigorously compute these spectral functions, a first-principles approach rooted in the Schwinger-Keldysh (or closed time path) formalism could be used. However, applying this formalism to our system is technically and computationally challenging, as it entails solving a coupled set of integro-differential equations for dynamical two- and four-point correlation functions. Moreover, while steps in this direction have been taken in the non-relativistic and dilute regime in, e.g., reference~\cite{Binder:2018znk}, a precise understanding of how to treat bound states in the relativistic regime remains missing. Therefore,
before committing ourselves to overcoming these non-trivial hurdles, we first use physical arguments to assess whether the expected effect on $\Delta \Neff$ justifies the effort.


Since the evolution is close to equilibrium and the electrons/positrons are mostly non-relativistic, one may (as also discussed in~\cite{Binder:2018znk}) recast the problem in terms of a finite-temperature effective potential $V(r)$, which in the hard thermal loop (HTL) limit takes the form~\cite{Laine:2006ns,Binder:2018znk} 
\begin{eqnarray}\label{VeffDef}
V(r) = -\alpha m_D - \frac{\alpha}{r}e^{-m_D r} - {\rm i} \alpha T \phi(m_D r),
\end{eqnarray}
with
\begin{align}\label{phidef}
    \phi(y) = 2\int_0^{+\infty} \d u \frac{u}{(u^2+1)^2} \left(1-\frac{\sin(uy)}{uy}\right),
\end{align}
where $m_D$ is the Debye mass, and $\alpha = e^2/(4\pi)$ the electromagnetic fine structure constant.
The HTL approximation assumes relativistic electrons ($m_e=0$) as well as soft momenta ($<\alpha T$) for the electrons, both of which assumptions are evidently inconsistent at $T \lesssim m_e$.  Nonetheless, physically it seems sensible to expect the functional form of \eqref{VeffDef} to hold also in the $T \lesssim m_e$ regime, although the Debye mass $m_D$ appearing in the expression may need to be modified from its standard HTL form $m_D^{\rm HTL}= e T/\sqrt{3}$.   Here, our strategy is to substitute $m_D$ with a value extracted from the {\it full} photon propagator without HTL approximation---to be elaborated on later. 

Inspecting the effective potential~\eqref{VeffDef}, the usual quasiparticle picture applies when the imaginary part  is small. In this limit, one may obtain the energies of bound states from ${\rm Re}[V(r)]$, while ${\rm Im}[V(r)]$ gives their thermal width. 
This understanding then allows us to factorise the problem of bound-state formation into two separate questions:
\begin{enumerate}[label=(\roman*)]
    \item At which temperatures can positronium exist in the early universe plasma?  
    \item If positronia are allowed to form, how fast do they equilibrate?
\end{enumerate}
We shall examine these two questions in the following subsections.


\subsection{Existence of the positronium at high temperatures}
\label{sec:melting}

It is well known that bound states should generally cease to exist at high temperatures---a phenomenon called ``melting''---because of 
\begin{enumerate}[label=(\roman*)]
    \item Debye screening of the mediator, and\label{it:Debye} 
    \item Scatterings between the mediator and particles in the plasma.\label{it:Scatter} 
\end{enumerate}
The two can be roughly associated with the real and imaginary parts of $V(r)$, respectively, in the quasiparticle limit where the latter is small. 
The precise theoretical calculation of the melting temperature is intricate \cite{Laine:2006ns,Kim:2016kxt}.
Here, we use simple criteria to quantify the impact of these two effects to obtain a first realistic estimate of the positronium melting temperature.

\paragraph{Debye screening of the photon.} 
At $T>0$ the photon develops a thermal mass from forward-scattering with electrons present in the early universe plasma.
Following the original argument by Matsui \& Satz~\cite{Matsui:1986dk}, 
i.e.,~neglecting for now ${\rm Im}[V(r)]$, we see immediately that this thermal mass modifies the electrostatic potential (second term in equation~\eqref{VeffDef}) to
\begin{align}\label{YukawaPotential}
    V(r) \sim \frac{e^{-r/a_D}}{r},
\end{align}
suppressing it exponentially at large distances $r \gtrsim a_D$, where $a_D \equiv 1/m_D$ is the photon's Debye screening length. 
As discussed earlier, while the shape of the effective potential~\eqref{VeffDef} has, strictly speaking, been obtained in the HTL limit, 
a general estimate of $a_D$  applicable also in the non-relativistic (i.e., $T \lesssim m_e$) regime 
can be extracted from the $00$-component of the retarded longitudinal photon self-energy, $\Pi^{R}_{L,00}$, via
\begin{align}
    a_D = \frac{1}{\sqrt{-\operatorname{Re}\Pi^{R}_{L,00}\left(q_0=0,|\mathbf{q}|\rightarrow 0\right)}},
\end{align}
and we note that the full 1PI-resummed retarded photon propagator has been computed in reference~\cite{Drewes:2024wbw}; see also \cite{Braaten:1993jw,Scherer:2024uui} for different analytical approximations.
Equivalently, $a_D$ can be obtained from the QED equation of state via the relation (see, e.g.,~\cite{Kapusta:2006pm,Bellac:2011kqa})
\begin{equation}
   \operatorname{Re}\Pi^{R}_{L,00}(q_0=0, \vert{\bf q}\vert\rightarrow 0)=-e^2\frac{\partial^2 P(\mu, T)}{\partial \mu^2},\label{eq:dPmu2}
\end{equation} 
where $P(\mu, T)$ is the pressure of the $e^+e^-\gamma$ system with chemical potential $\mu$, given in the ideal gas limit by
\begin{equation}
\label{eq:pressureformula}
    P^{(0)}(\mu, T)=\frac{T}{\pi^2}\int_0^{\infty}{\rm d}pp^2\log\left[\frac{(1+e^{-(E_e-\mu)/T})^2}{1+e^{-p/T}}\right],
\end{equation} 
with $E_e=\sqrt{p^2+m^2_e}$.  Evaluating \eqref{eq:dPmu2} in the large $T$ limit ($m_e\rightarrow 0$) yields the well-known HTL result $\operatorname{Re}\Pi^{R, \rm HTL}_{L,00}=-e^2T^2/3$ \cite{Bellac:2011kqa} and hence $a_D^{\rm HTL}=\sqrt{3}/(eT)$.

As a first estimate of the positronium melting temperature, we follow the arguments presented in, e.g., \cite{Kim:2016kxt,Matsui:1986dk} and assume that melting occurs once the Debye screening length $a_D$ drops below the most likely radius of the positronium, $r_\ell \approx a_0 n^2$, where
\begin{align}
    a_0 = \frac{1}{m_{\rm r} \alpha} = \frac{8\pi}{m_e e^2}\simeq 536 ~\text{MeV}^{-1}
    \label{eq:Bohrradius}
\end{align}
is the positronium's Bohr radius, and $m_{\rm r}=m_e/2$ is the system's reduced mass.
Note that $a_0$ does not depend on the spin (singlet versus triplet) of the positronium configuration, nor is the running of $\alpha$ significant in the temperature range of interest.  We do caution however that where Debye screening of the electrostatic potential is effective at high temperatures, the Bohr radius as given in equation~\eqref{eq:Bohrradius} might not be a fully accurate way to assess the size of the positronium; for a first estimate, we shall neglect this detail.

\begin{figure}
    \centering
    \includegraphics[width=.7\textwidth]{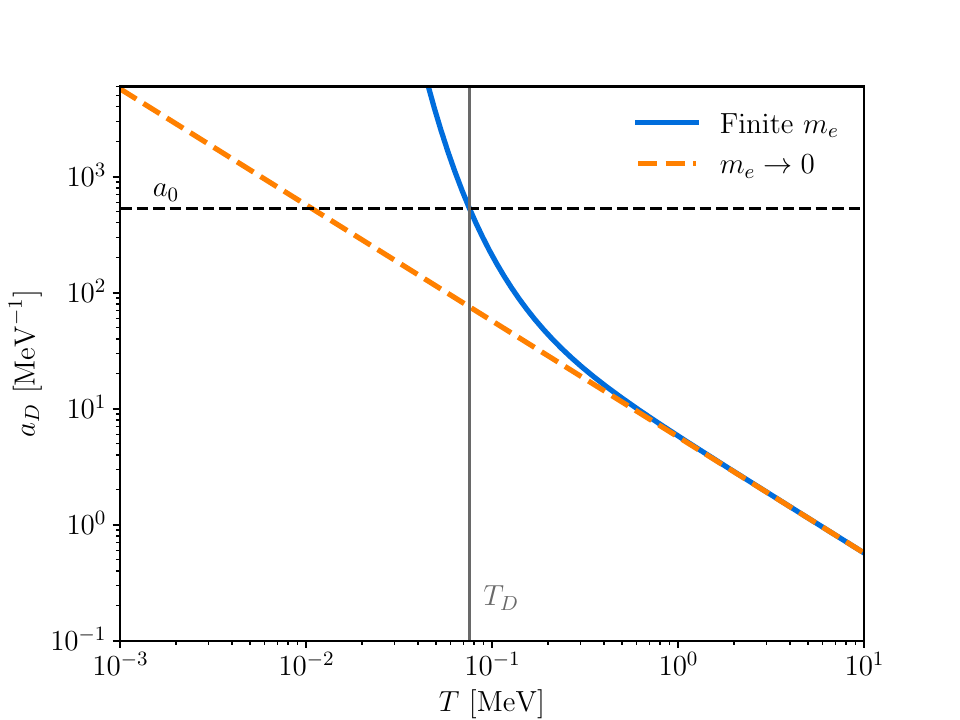}
    \caption{Debye screening length as a function of the temperature computed from the 1PI-resummed photon propagator (blue) and the HTL approximation (orange). Observe that while the two estimates coincide at $T \gtrsim m_e$, the former is significantly larger at $T \lesssim m_e$.
    We compare $a_D$ to the Bohr radius $a_0$ of the positronium state in vacuum (black dashed horizontal line). The vertical line represents $T_{D}$, the temperature at which $a_0 = a_D$ and Debye screening starts to be effective at $T>T_{D}$.}
    \label{fig:meltingboundstates}
\end{figure}

Figure~\ref{fig:meltingboundstates} shows the Debye screening length $a_D$ as a function of temperature,  extracted from the 1PI-resummed photon self-energy~\cite{Drewes:2024wbw} (or, equivalently, from equation~\eqref{eq:dPmu2}), assuming a vanishing chemical potential $\mu = 0$ for the electrons.  For completeness, we also plot the HTL approximation $a_D^{\rm HTL}$.  Applying the criterion that bound states only exist when%
\footnote{For a standard Yukawa potential, numerical analyses~\cite{sachs1938calculations,rogers1970bound,Matsui:1986dk} of the eigenstates of the Schrödinger equation found a slightly more constraining condition, $a_D \gtrsim a_0/0.84$, for $n=1$. Applying this condition to our case shifts the melting temperature to $72$ keV. We shall however neglect this effect for a first estimate, as the non-trivial momentum dependence of the 1PI-resummed photon propagator might in any case spoil this numerical estimate based on the Yukawa potential. \label{footnote:numericalSchrodinger}} 
$a_D> r_\ell$, we see immediately that $n=1$ positronium formation may be possible at temperatures 
\begin{align}
\label{eq:Debyemeltingresult}
    T \lesssim T_{D}\simeq 75.6~\mbox{keV},
\end{align}
while the $n=2$ state has a melting temperature of $55$ keV.  Temperatures exceeding the constraint~\eqref{eq:Debyemeltingresult} are marked in figure~\ref{fig:tanh} by a grey-shaded region; no positronium exists at these temperatures within this approximation.  Where positronium formation is possible, the corresponding change in $\Neff^{\rm SM}$ can be up to $|\Delta \Neff| \simeq 10^{-4}$ assuming instantaneous equilibration (see figure~\ref{fig:tanh}).  Non-instantaneous equilibration generally results in a larger effect.  Using for example the kernel $\sigma(T)=1-\exp[(a_D/a_0)^k]$, mimicking a progressive positronium production around $T_D$, yields $\Delta N_{\rm eff}=-2.25\times 10^{-3}, -2.61\times 10^{-4}$ for $k=2,4$ respectively. 

Note that had we used the HTL approximation $a_D^{\rm HTL}$, we would have arrived at a melting temperature of $10.7$~keV---much lower than the estimate~\eqref{eq:Debyemeltingresult} and
much too low to impact $\Neff^{\rm SM}$ in an appreciable way. 
Physically, the higher temperature~\eqref{eq:Debyemeltingresult} can be understood as a result of the Boltzmann-suppressed densities of real electrons and positrons at $T \lesssim m_e$ (such suppression is absent in the HTL limit), leading to less screening. 
However, even substituting the more correct estimate~\eqref{eq:dPmu2} of $a_D$ for $a_D^{\rm HTL}$ may still fail to capture the highest temperature~\eqref{eq:Debyemeltingresult} at which bound states can exist for the following reasons. Firstly, in addition to the Yukawa-like term~\eqref{YukawaPotential}, the real part of the full potential~\eqref{VeffDef} contains the so-called Salpeter term, $-\alpha m_D$, which directly impacts the bound state's spectral function. Secondly, the imaginary part of $V(r)$ \cite{Laine:2006ns,Brambilla:2008cx,Vairo:2008hzj,Escobedo:2008sy,Biondini:2023zcz}, which was neglected in the potential \eqref{YukawaPotential}, implies that the transition to the regime in which bound states can exist is not a sharp one, as discussed in the following.
Lastly, the non-trivial momentum dependence of the 1PI-resummed photon propagator could generate sizeable deviations from the potential \eqref{VeffDef}, obtained in the HTL limit; see also footnote \ref{footnote:numericalSchrodinger}.

\paragraph{Dissociation from scatterings:} Beyond Debye screening of the electric field, dissociation can also be driven by the photon's real scatterings
with the rest of the QED bath~\cite{Kim:2016kxt,Laine:2006ns}, an effect often referred to as Landau damping.
Scattering has even been shown to be the dominant melting mechanism in the case of heavy quarkonia~\cite{Dominguez:2008be,Brambilla:2008cx,Vairo:2008hzj}. 
In terms of the effective potential~\eqref{VeffDef}, dissociation due to scatterings can be ascribed to ${\rm Im}[V(r)]$,%
\footnote{Generally, the real and imaginary corrections to the photon propagator can be related respectively to the real and imaginary parts of the refractive index, with the latter having its physical origin in scatterings with particles in the medium.}
which gives the bound states a finite width.
As the temperature increases, this width increases, progressively bridging the gap between the bound state and the continuum.  When the thermal width is much larger than the positronium's binding energy, the bound-state contribution to the spectral function is not well-separated from the continuum, and the positronium is expected to dissociate into an electron-positron pair.

To estimate this thermal width~$\Gamma$, we use the results from reference~\cite{Laine:2006ns} and approximate 
\begin{align}
\label{eq:widthstaticpotential}
    \Gamma =  \alpha T \int \d^3{\bf r} \; \phi(r/a_D) |\Psi^{\rm Ps}_{1s}({\bf r})|^2,
\end{align}
where $\Psi^{\rm Ps}_{1s}$ represents the ground state of positronium, and $\phi(r/a_D)$, with $r=|\mathbf{r}|$, has been defined in equation~\eqref{phidef}.
Again, we note that, strictly speaking, both equations~\eqref{phidef} and~\eqref{eq:widthstaticpotential} have been derived using the HTL form of the photon propagator.  But, as with our estimate of~$T_D$, we take equation~\eqref{eq:widthstaticpotential} at face value and evaluate it using the Debye screening length~$a_D$ derived from the full 1PI-resummed photon propagator. 
Then, as a simple criterion to determine the dissociation temperature $T_{\rm dis}$, we define it to be the temperature at which  
the width $\Gamma$ equals the binding energy $\Delta E_n$, i.e., 
\begin{align}\label{ScatterinfTdis}
    \Gamma(T_{\rm dis}) =  \Delta E_n = \frac{m_e \alpha^2}{4 n^2}.
\end{align}
Assuming the positronium wavefunction takes its zero-temperature form (a reasonable approximation when $a_D \gg a_0$ is satisfied), we find from evaluating~\eqref{eq:widthstaticpotential} for the $n=1$ state
\begin{equation}
T_{\rm dis} \simeq 45.8~{\rm keV},
\label{eq:tdis}
\end{equation}
well below the melting temperature $T_D \simeq 75.6$~keV from the Debye screening argument. 
(Again, we do not consider excited states, as the same argument forbids their production to begin before, e.g., $T\simeq 40.5$~keV for $n=2$.)
The dissociation temperatures of the hydrogen and the muonic hydrogen atoms have been computed in~\cite{Escobedo:2008sy,Escobedo:2010tu} using a similar approach but with a different {\it ansatz} for the  resummed photon propagator; we have checked that our method reproduces the results of~\cite{Escobedo:2008sy,Escobedo:2010tu} to within $10\%$ for these two systems.

The constraint from scatterings is denoted by the hatched region in figure~\ref{fig:tanh}. Taken at face value along with the assumption of instantaneous positronium equilibration, such a low formation temperature as indicated by~\eqref{eq:tdis} would yield a negligible change in $\Neff^{\rm SM}$ of $\Delta \Neff \simeq - 3\times 10^{-8}$. However, we caution that there are many uncertainties in this calculation at this stage. To begin with, because $T_{\rm dis} < T_D$, there is a temperature window in which the quasiparticle picture does not hold and we cannot write down an on-shell Boltzmann equation to directly evaluate the positronium equilibration temperature and rate. 
Nonetheless, dissociation due to scatterings is expected to be a more continuous process arising from the progressive increase of the thermal width $\Gamma$ with temperature, in contrast to melting from Debye screening where a distinct (though temperature-dependent) mediator mass sets the scale above which bound states cannot exist.%
\footnote{We have reached this conclusion via a very simple understanding of the thermal positronium spectral function, namely, by comparing the separation between the positronium ground state energy in vacuum and the continuum of scattering states with energy $E>0$,
to the  estimate~\eqref{eq:widthstaticpotential} of the positronium's thermal width. In reality, the positronium spectral function at $T>0$ is continuous (and non-zero) also for $E<0$, and the position of the quasiparticle peaks changes with $T$. While integrals over this spectral function 
(as they appear in collision terms)
tend to be dominated by the pronounced quasiparticle peaks when they exist, sum rules ensure that there are always non-zero contributions from the melted peak for both $E<0$ and $E>0$ for $T>0$. This suggests a gradual increase of the impact of bound states in the $T_{\rm dis}$-to-$T_D$ window (and likely even at $T>T_D$, implying that even $T_D$ should strictly speaking not be viewed as a sharp upper limit).}  
 In terms of its impact on $\Delta \Neff$, a reasonable expectation is that dissociation will lead to a slower approach to equilibrium in the hatched region than is suggested by the instantaneous equilibration approximation, an effect that may to an extent be captured by a larger $\Delta T/T_{\rm Ps}^{\rm eq}$ in our phenomenological modelling  of positronium equilibration.\footnote{Because the quasiparticle picture/standard Boltzmann equations do not apply in the 
 transition region, it is not straightforward to relate $\Delta T$ to the microphysical interaction rates.} %
 This would in turn drive up $\Delta \Neff$ from the instantaneous equilibration estimate.  Modelling the effect precisely is however beyond the scope of this work,
as an accurate description would require that we study the form of the positronium spectral functions, such as in, e.g.,~\cite{Binder:2018znk}, which we leave for future work.


\subsection{Positronium equilibration rate}
\label{sec:poseqtemperature}

Having established an estimate of the upper limit on the positronium formation temperature, namely, $T_D \simeq 80$~keV, and supposing that Boltzmann equations provide a valid description of positronium production, we now turn to the question of whether this formation can happen quickly enough to produce a sizeable impact on $\Neff^{\rm SM}$. Properties of positronium have been under scrutiny for decades; see, e.g., \cite{Stroscio:1975fa} and references therein.  Positronium-like  dark matter bound states have also been the subject of intensive study because of their potential impact on dark matter freeze-out and relic abundance as well as on indirect dark matter searches~(e.g.,~\cite{Pospelov:2008jd,March-Russell:2008klu,Shepherd:2009sa,vonHarling:2014kha}). The Boltzmann equations describing the evolution of  positronium-like states and the associated production rates in a non-relativistic plasma are very well known in the literature (e.g., \cite{vonHarling:2014kha,Vasilaki:2024fph}).

In the situation under consideration, given that electrons remain in kinetic equilibrium in the temperature range relevant for $\Neff^{\rm SM}$, the Boltzmann equation for the positronium number density $n_{\rm Ps}$ takes the form
\begin{align}
    \label{eq:Positroniumevolutiongeneral}
     \frac{\d}{\d x}(n_{\mathrm{Ps}}x^3) &= -\frac{1}{Hx}\left[\Gamma^{\mathrm{dec}} +\Gamma^{\mathrm{ion}}(T)\right]\left(n_{\mathrm{Ps}}x^3-n_{\mathrm{Ps}}^{\mathrm{eq}}x^3\right),
\end{align} 
where  $n_{\mathrm{Ps}}^{\mathrm{eq}}$ is the equilibrium positronium number density.  In the non-relativistic regime, the ionisation rate $\Gamma^{\mathrm{ion}}$ and the decay rate $\Gamma^{\mathrm{dec}}$ are given  by~\cite{Vasilaki:2024fph}
\begin{subequations}
\label{eq:Boundstaterates}
\begin{align}
    \Gamma^{\mathrm{ion}}(T) &= \frac{m_{\rm r} \alpha^5}{8\pi}\int_0^{+\infty} \frac{\d \xi\; \xi^{-4}}{\mathrm{exp}\left[\left(\frac{1}{n^2}+\frac{1}{\xi^2}\right)\frac{m_e\alpha^2}{4T}\right]-1}S_{\mathrm{Ps}}^{\rm BSF}(\xi),\\
    \Gamma^{\mathrm{dec}} &=
\begin{cases}
m_{\rm r} \alpha^5, & \text{for } s=0, \\
\frac{4(\pi^2-9)}{9\pi}m_{\rm r} \alpha^6, & \text{for } s=1,
\end{cases}
\end{align}
\end{subequations}
where $s=0,1$ denote, respectively, the singlet and triplet positronium states, and
\begin{align}
    S_{\mathrm{Ps}}^{\rm BSF}(\xi) = \frac{2^9}{3} \frac{\xi^4 e^{-4 \xi \mathrm{cot}^{-1}\left(1/\xi\right)}}{(\xi^2+1)^2} \frac{2\pi\xi}{1-e^{-2\pi\xi}} .
\end{align}
To establish the positronium abundance as a function of temperature $T$, one needs in principle to solve two Boltzmann equations~\eqref{eq:Positroniumevolutiongeneral} using the rates~\eqref{eq:Boundstaterates} to keep track of the $s=0$ and $s=1$ states separately.   However, even without an explicit numerical solution, it is straightforward to verify that, at $T \simeq 80$ keV, the decay and ionisation rates are well in excess of the Hubble expansion rate, i.e.,
\begin{align}
    \left.\frac{\Gamma^{\mathrm{dec}} +\Gamma^{\mathrm{ion}}}{H}\right|_{T=T_D} \simeq  10^{16} \gg 1.
\end{align} 
 This indicates that, {\it in the absence of the melting and dissociation effects} discussed in section~\ref{sec:melting}, the equilibration of positronium would be instantaneous.

Finally, we remark again that the rates~\eqref{eq:Boundstaterates} apply in the non-relativistic regime only. While this should be a good approximation at $T \lesssim 80$ keV, one may wonder if it overestimates contributions from relativistic electrons and positrons if blindly applied to a thermal distribution. We therefore consider also a worst-case scenario estimate for the rates by (i)~neglecting the decay contribution, and, for the ionisation rate, (ii)~selecting only electron momenta corresponding to kinetic energies smaller than the positronium binding energy, i.e.,
\begin{equation}
\begin{aligned}
\label{eq:rateextrapolation}
\Gamma_{\rm extr} & = \Gamma^{\mathrm{ion}} \left(\frac{n_e^{\mathrm{eq}}(p<p_\mathrm{lim})}{n_e^{\mathrm{eq}}}\right)^2\\ 
    &= \Gamma^{\mathrm{ion}} \left(\frac{\int_{0}^{p_\mathrm{lim}}p^2f_{\rm FD}(p/T,m_e/T) \d p}{\int_{0}^{+\infty} p^2f_{\rm FD}(p/T,m_e/T) \d p}\right)^2,
\end{aligned}
\end{equation}
where $p_{\mathrm{lim}}$ is the largest momentum for which the electron-positron relative velocity $v_{\rm rel}$ satisfies $v_{\mathrm{rel}}<\alpha$. But, even with this extremely conservative estimate of the production rate $\Gamma_{\rm extr}$, we still find   $\Gamma_{\rm extr}/H \simeq 3\times 10^2$ at $T \lesssim 80$~keV,
indicating again a rapid  approach to equilibrium as soon as formation is allowed.  We therefore conclude that the formation of positronium is not limited by the production rates, but only by the behaviour of the system  between $T_D$ and $T_{\rm dis}$.  Since the overall change in $\Neff^{\rm SM}$ in this region is exponentially sensitive to the positronium equilibration temperature, further investigations will be necessary to characterise with certainty the ultimate impact of positronium formation on $\Neff^{\rm SM}$.


\section{Positronium as non-ideal gas correction}
\label{sec:nonideal}

 Up to now we have considered a scenario in which the production of positronium at $T_{\rm Ps}^{\rm eq} \lesssim 80$~keV is accompanied by a net increase in the total entropy of the QED plasma at production time.  The corresponding correction to $N_{\rm eff}^{\rm SM}$ is directly linked to this increase; see, e.g., equation~\eqref{eq:dNeff}.
 Having established in section~\ref{sec:poseqtemperature}, however, that the positronium equilibration rate is in fact very large, it is also useful to consider the opposite limit in which the appearance of bound states at $T_{\rm Ps}^{\rm eq}$ does not generate entropy in the QED sector {\it overall}. That is, the QED sector remains in equilibrium within itself at all times after neutrino decoupling.
In this limit, any changes to $N_{\rm eff}^{\rm SM}$ due to bound states will depend solely on their contribution to the entropy of the QED plasma {\it at neutrino decoupling}, which,  in the instantaneous-decoupling approximation, can be established from non-ideal corrections to the QED equation of state at $T=T_d \simeq 1.3$~MeV.

Given that $T_d \gg T_{\rm Ps}^{\rm eq}$, one might naively conclude that the absence of bound states at $T \simeq T_d$ must incur a negligible effect on $N_{\rm eff}^{\rm SM}$ in this scenario. This simplified picture may however be misleading for the following reason. The same long-range electron-positron interaction that gives rise to bound states at low temperatures also implies that resonant interactions amongst the scattering-state electrons and positrons are possible prior to bound-state formation.  These non-perturbative effects alter the QED equation of state (EoS) away from the ideal-gas behaviour and could persist at some level even at much higher temperatures, e.g., around $T= T_d$, thereby impacting on $N_{\rm eff}^{\rm SM}$.
Indeed, as we shall show below, in this entropy-conserving scenario, the entropy residing in bound states---once they are allowed to form---can be envisaged as having been ``stolen'' from the resonant $e^+e^-$-scattering states, whose own entropy decreases accordingly.

To estimate $\Delta N_{\rm eff}$ in this scenario quantitatively, we note first of all that no fully non-perturbative EoS formula exists in the \emph{relativistic} and non-dilute regime, i.e., $T \gg m_e,\mu_e$, with $\mu_e$ the electron chemical potential.  However, the \emph{non-relativistic} limit $T \ll m_e$ is well described by the so-called Beth-Uhlenbeck formula~\cite{Beth:1937zz}. This formula consistently incorporates both bound and interacting scattering states in the non-ideal pressure. Importantly, as a fully non-perturbative formula, Beth-Uhlenbeck contains also perturbative contributions to the QED EoS in the non-relativistic limit that, in the context of $N_{\rm eff}^{\rm SM}$, have already been accounted for elsewhere (e.g.,~\cite{Bennett:2019ewm}), and care needs to be taken to avoid double counting.

Our strategy in this section, therefore, is as follows.  We first isolate in section~\ref{sec:eq} the leading-order perturbative correction to the EoS from a non-relativistic partition function that forms the basis of the Beth-Uhlenbeck formula.  Comparison of this leading correction with the fully non-perturbative formula in section~\ref{sec:BU} then allows us to demonstrate that the system remains perturbative even at the time of bound-state formation and hence provides a means for us to consistently extract the higher-order corrections---including non-perturbative corrections from bound states---contained in the formula. Non-technical readers wishing to skip these details may jump directly to section~\ref{sec:Neff}, where we extrapolate the extraction approach to $T=T_d$ to obtain an estimate of (or, more correctly, an upper limit on) $\Delta N_{\rm eff}$ due to bound-state formation in this entropy-conserving scenario.


\subsection{Perturbative computation of non-ideal corrections}
\label{sec:eq}

The goal of sections~\ref{sec:eq} and~\ref{sec:BU} is to demonstrate that perturbation theory can be applied to the system at hand in a controlled way, even at times when bound states begin to form.  To this end, our first task is to determine the leading-order non-ideal correction in the coupling~$e$ to the QED EoS in the non-relativistic limit.  We begin by noting for future reference that, 
in equilibrium, the energy density~$\rho$ and the entropy~$s$ are related to the pressure density via
\begin{align}
P = \frac{T}{{\cal V } } \ln Z \;, \quad \rho = \frac{T^2}{{\cal V } } \partial_T \ln Z = - P + T \partial_T P \;, \quad s = \frac{1}{{\cal V } } \partial_T (T \ln Z ) = \frac{\rho + P}{T} , \label{eq:entr}
\end{align}
where ${\cal V }$ is the spatial volume, $Z=\text{Tr} [ e^{-\beta H}]$ the partition with Hamiltonian $H=H_0 + H_\text{int}$, $\beta = 1/T$, and we have set the chemical potentials to zero here for simplicity. 

In {\it relativistic} QED, the standard textbook procedure is to treat interaction term $\mathcal{L}_{\text{int}} = -e \bar{\psi}\gamma^\mu \psi A_\mu$ as a perturbation~\cite{Kapusta:2006pm}, and expand the partition function  in powers of the QED coupling $e$ 
\begin{align}
\ln Z = \ln Z^{(0)} + \ln Z^{(2)} + \ldots,
\end{align}
where $\ln Z^{(2)}$ constitutes  the first, i.e., ${\cal O}(e^2)$, non-ideal correction.
The corresponding ${\cal O}(e^2)$ correction to the pressure, $P^{(2)}$, can be written as a sum over three parts, $P^{(2)}=P^{(2a)}+P^{(2b)}+P^{(2c)}$, where the individual contribution reads
\begin{subequations}
\label{eq:p2}
\begin{align}
P^{(2a)} &= \frac{e^2 m^2_e}{4 \pi^4} \int \int \text{d} p\;  \text{d} \Tilde{p} \; \frac{p \Tilde{p}}{E_p E_{\Tilde{p}}} \ln \bigg| \frac{p+\Tilde{p}}{p-\Tilde{p}} \bigg| f_F(E_p) f_F(E_{\tilde{p}}) \stackrel{\rm NR}{\simeq}  \frac{e^2 m^2_e T^2}{8 \pi^3} e^{-\beta 2 m_e}\;, \label{eq:2a}\\
P^{(2b)} &= - \frac{e^2}{2 \pi^4} \left( \int \text{d} p \; \frac{p^2}{E_p} f_F(E_p)\right)^2 \stackrel{\rm NR}{\simeq} - \frac{e^2 m_e T^3}{4 \pi^3} e^{-\beta 2 m_e}  \;,\\
P^{(2c)} &= - \frac{e^2 T^2}{6 \pi^2} \int \text{d} p \;\frac{p^2}{E_p} f_F(E_p) \stackrel{\rm NR}{\simeq} - \frac{e^2}{3} \left(\frac{m_e T^7}{(2\pi)^3}\right)^{1/2} e^{-\beta m_e} \;,
\end{align}
\end{subequations}
with $f_F(E) = 1/(e^{E/T}+1)$ denoting the standard Fermi-Dirac distribution. For later reference, we have evaluated each contribution in the non-relativistic limit; in this limit, $P^{(2a)}$ is the leading correction in the squared number density $n_e^2 \sim m_e^3 T^3 e^{-2 \beta m_e}$, while $P^{(2c)}$ is linear in~$n_e$.

The pressure correction in the non-relativistic limit can also be obtained directly from the non-relativistic interaction Hamiltonian
\begin{equation}
\begin{aligned}
H^\text{int}_\text{NR}= - \frac{1}{2} \sum_{\sigma,\sigma^\prime} \int \text{d}^3\mathbf{x} \text{d}^3
\mathbf{y} V(|\mathbf{x}-\mathbf{y}|) \bigg[ &\eta^{\dagger}_\sigma(\mathbf{x})  \eta^{\dagger}_{\sigma^\prime}(\mathbf{y})   \eta_{\sigma^\prime}(\mathbf{y}) \eta_{\sigma}(\mathbf{x}) \\ +   & \xi_{\sigma}(\mathbf{x}) \xi_{\sigma^\prime}(\mathbf{y})  \xi^{\dagger}_{\sigma^\prime}(\mathbf{y}) \xi^{\dagger}_\sigma(\mathbf{x})  \\ - 2 &\eta^{\dagger}_\sigma(\mathbf{x}) \xi_{\sigma^\prime}(\mathbf{y})  \xi^{\dagger}_{\sigma^\prime}(\mathbf{y})\eta_{\sigma}(\mathbf{x})   \bigg] ,
 \label{eq:ham}
\end{aligned}
\end{equation}
where $V$ is the potential, the field operator $\eta^\dagger$ creates an electron and $\xi$ a positron, and summation is over the spin indices $\sigma, \sigma'$.  
Performing a real-time computation yields an ${\cal O}(e^2)$ non-relativistic correction to the pressure given by
\begin{align}
P^{(2)}_{\text{NR}}=\frac{T}{{\cal V } } \ln Z^{(2)}_\text{NR} = \frac{1}{{\cal V } }\frac{1}{2} \sum_{\sigma \sigma^\prime} \int \text{d}^3\mathbf{x} \text{d}^3
\mathbf{y} V(|\mathbf{x}-\mathbf{y}|) \bigg[ G^{++--}_{\eta \eta,0}(t\mathbf{x},\mathbf{y},t\mathbf{x},\mathbf{y}) + G^{++--}_{\xi \xi,0} -2 G^{++--}_{\eta \xi,0}  \bigg], \label{eq:p2nr}
\end{align}
where the four-point equal-time correlators are in normal order according to \eqref{eq:ham}, e.g.,
\begin{align}
G^{++--}_{\eta \eta,0}(t\mathbf{x},\mathbf{y},t\mathbf{x},\mathbf{y})&\equiv \langle \eta^{\dagger}_\sigma(\mathbf{x})  \eta^{\dagger}_{\sigma^\prime}(\mathbf{y})   \eta_{\sigma^\prime}(\mathbf{y}) \eta_{\sigma}(\mathbf{x}) \rangle_0 \,,
\end{align}
and $\langle \dots \rangle_0$ denotes a thermal average with respect to the free Hamiltonian $H_0$. 

\begin{figure}
    \centering
    \includegraphics[width=0.8\linewidth]{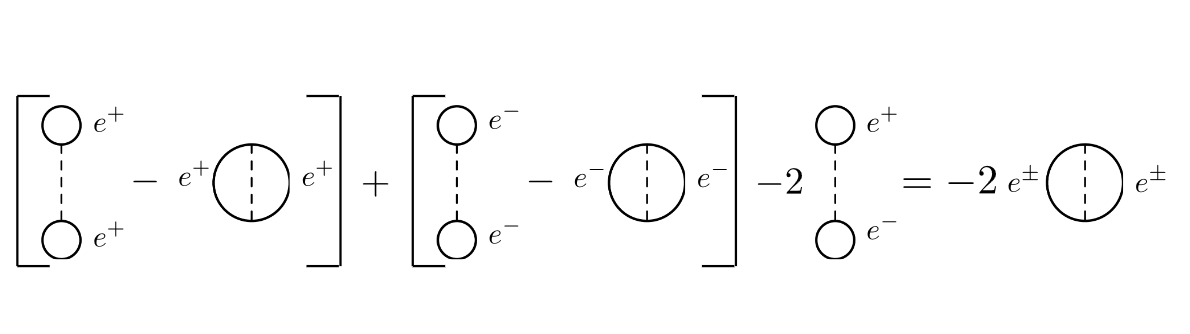}
    \caption{Feynman diagrammatic representation of equation~\eqref{eq:p2nr}. In a symmetric plasma, the ``dumbbells'' diagrams cancel, leaving us with the ``oyster'' diagram to the right of the equal sign.}
    \label{fig:epem}
\end{figure}

Using the equations of motion~\eqref{eq:eom1} and~\eqref{eq:eom2} in integral form, the four-point correlators can be expressed in terms of free two-point functions as
\begin{equation}
\begin{aligned}
G^{++--}_{\eta \eta,0}(t\mathbf{x},\mathbf{y},t\mathbf{x},\mathbf{y}) &=  G_{\eta,0}^{+-}(t \mathbf{x}, t \mathbf{x} ) G_{\eta,0}^{+-}(t \mathbf{y}, t \mathbf{y} ) - G_{\eta,0}^{+-}(t \mathbf{x}, t \mathbf{y} ) G_{\eta,0}^{+-}(t \mathbf{y}, t \mathbf{x} ), \label{eq:etaeta} \\
G^{++--}_{\xi \xi,0}(t\mathbf{x},\mathbf{y},t\mathbf{x},\mathbf{y}) &=  G_{\xi,0}^{+-}(t \mathbf{x}, t \mathbf{x} ) G_{\xi,0}^{+-}(t \mathbf{y}, t \mathbf{y} ) - G_{\xi,0}^{+-}(t \mathbf{x}, t \mathbf{y} ) G_{\xi,0}^{+-}(t \mathbf{y}, t \mathbf{x} ),  \\
G^{++--}_{\eta \xi,0}(t\mathbf{x},\mathbf{y},t\mathbf{x},\mathbf{y}) &=  G_{\eta,0}^{+-}(t \mathbf{x}, t \mathbf{x} ) G_{\xi,0}^{+-}(t \mathbf{y}, t \mathbf{y} ) ,
\end{aligned}
\end{equation}
where note that the degenerate cases (i.e., the $\eta \eta$ and $\xi \xi$ correlators) have two contributions each. Inserting~\eqref{eq:etaeta} into~\eqref{eq:p2nr}, figure~\ref{fig:epem} shows the diagrammatic representation of the ${\cal O}(e^2)$ non-relativistic correction to the pressure. The ``dumbbell'' diagrams correspond to the Hartree corrections, which cancel in a symmetric plasma, i.e., where the electron and positron number densities are equal, such that $G_{\eta,0}^{+-}(t \mathbf{x}, t \mathbf{x} ) = G_{\xi,0}^{+-}(t \mathbf{x}, t \mathbf{x})$. The remainder is the exchange or ``oyster'' diagram, also known as the Hartree-Fock correction, which evaluates using Wigner coordinates to
\begin{equation}
\begin{aligned}
P^{(2)}_{\text{NR}} &=  (-2) \times \frac{1}{2}  \sum_{\sigma \sigma^\prime} \int \text{d}^3 \mathbf{r}\;  V(r)   G_{\eta,0}^{+-}(0, \mathbf{r}) G_{\eta,0}^{+-}(0, - \mathbf{r})  \\
& = -  2 \int \frac{\text{d}^3 \mathbf{p}_1}{(2 \pi)^3} \frac{\text{d}^3 \mathbf{p}_2}{(2 \pi)^3} V(\mathbf{p}_1- \mathbf{p}_2) e^{- \beta (2 m_e + \mathbf{p}_1^2/(2m_e) + \mathbf{p}_2^2/(2m_e)) }  \\
&= \frac{e^2 m^2_e T^2}{8 \pi^3} e^{-\beta 2 m_e}.
\label{eq:p2NR}
\end{aligned}
\end{equation}
Here, we have used the real-time Feynman rules in momentum space for the free non-relativistic two-point functions in equilibrium provided in appendix~\ref{app:FR}, and the Coulomb potential in momentum space is $V(\mathbf{q})=-e^2/\mathbf{q}^2$. Comparing this result~\eqref{eq:p2NR} with equation~\eqref{eq:p2}, we see immediately that $P^{(2)}_{\text{NR}}$ coincides with $P^{(2a)}$ in the non-relativistic limit, while the sub-leading term $P^{(2b)}$ and the correction $ P^{(2c)}$ linear in $n_e \sim e^{-\beta m_e}$ are absent.

Pressure corrections proportional to $n_e^2 \sim e^{-\beta 2 m_e}$ correspond to the \emph{second} virial coefficient in a virial expansion, and the normal-ordered Hamiltonian~\eqref{eq:ham} is commonly used in the literature~\cite{kremp2005quantum, bookEB} to compute this quantity.  Indeed, for our purpose of estimating the effects of bound states on the QED EoS, the second virial coefficient is the quantity of interest, as it accounts for bound states  when computed non-perturbatively. These considerations lead us to conclude that, for a fixed-order perturbation theory in $e$ to be under control, higher-order/non-perturbative corrections (due to, e.g., bound states) to $P_{\rm NR}$ must remain small relative to $P^{(2)}_{\text{NR}}$ computed perturbatively in equation~\eqref{eq:p2NR}.   We turn to the non-perturbative evaluation of $P_{\rm NR}$ below.

\subsection{Beth-Uhlenbeck formula}
\label{sec:BU}

The Beth-Uhlenbeck formula~\cite{Beth:1937zz} is a non-perturbative expression for the non-relativistic pressure that  accounts for both bound and interacting scattering states, and can be derived from the partition function of the non-relativistic Hamiltonian~\eqref{eq:ham}; see, e.g.,~\cite{kremp2005quantum, bookEB} and appendix~\ref{app:bu}. The formula takes the form
\begin{align}
\beta(P_\text{NR}-P^{(0)}_\text{NR}) =   4 \left(\frac{2m_e T }{2\pi}\right)^{3/2} e^{-\beta 2 m_e } \sum_{\ell=0}^\infty (2\ell +1) b_\ell \;, \label{eq:BU}
\end{align}
with
\begin{align}
 b_\ell = \sum_{n \geq \ell +1} e^{-\beta E_{n\ell}} + \frac{1}{\pi} \int_0^\infty \text{d} E \; e^{-\beta E} \frac{\text{d} }{\text{d}E} \left\{ \delta_\ell^{e^+ e^-}+ \left(\frac{1}{2} - \frac{(-1)^\ell}{4}\right) \left( \delta_\ell^{e^+ e^+} +  \delta_\ell^{e^- e^-} \right) \right\} \;, \label{eq:BU2}
\end{align}
where $E_{n\ell}$ is the binding energy of the positronium  bound state with principle quantum number $n$ and angular momentum $\ell$, and $\delta_\ell$ denotes the quantum mechanical phase shift of the scattering states, with $\delta_\ell^{e^+ e^+} = \delta_\ell^{e^- e^-}$ for the repulsive contributions.

We proceed by first applying perturbation theory to~\eqref{eq:BU} to identify the leading correction~\eqref{eq:p2NR}.
 In the Born approximation of \eqref{eq:BU2}, the first, bound-state term is zero, while the phases in the integral are given by
\begin{align}
\delta_{\ell,\text{Born}}^{e^+ e^-/e^+ e^+} = \mp \frac{m_e}{k} \int_0^\infty \text{d}r \; V(r) \big[ k r j_\ell(kr) \big]^2 \;, \label{eq:phaseborn}
\end{align}
with $k=\sqrt{m_e E}$ and $j_\ell(kr)$ is a spherical Bessel function of degree~$\ell$.   Observe that the phases $\delta_{\ell,\text{Born}}^{e^+ e^-}$ and $\delta_{\ell,\text{Born}}^{e^+ e^+}$ differ only by a sign. This means that $\delta_{\ell,\text{Born}}^{e^+ e^-}$ in equation~\eqref{eq:BU2}  cancels entirely against $\left(\delta_{\ell,\text{Born}}^{e^+ e^+}+\delta_{\ell,\text{Born}}^{e^- e^-}\right)/2$, resulting in
\begin{align}
\sum_{\ell=0}^\infty (2\ell +1) b_\ell^{\text{Born}} = - \frac{1}{2} \frac{1}{\pi} \int_0^\infty \text{d} E e^{-\beta E} \frac{\text{d} }{\text{d}E} \underbrace{\sum_{\ell=0}^\infty (2\ell +1) (-1)^\ell \delta^{e^+ e^+}_{\ell, \text{Born}}}_{= k f_\text{Born}^{e^+ e^+}(\theta=\pm  \pi)}, \label{eq:sumborn}
\end{align}
where the sum over all partial waves can be identified with the scattering amplitude $f_\text{Born}^{e^+ e^+}(\theta=\pm  \pi)$. For a Yukawa potential with screening mass $m_D$, the scattering amplitude is given by
\begin{align}
f_\text{Born, Yukawa}^{e^+ e^-/ e^+ e^+}(\theta) &= \pm  \frac{\alpha m_e}{m_D^2+ 4 k^2 \sin^2(\theta/2)} \;,
\label{eq:bornyukawa}
\end{align}
and is not divergent at $\theta=\pm \pi$ in the Coulomb (i.e., $m_D \rightarrow 0$) limit.  Then, the energy integral in \eqref{eq:sumborn} can be performed analytically. Taking the Coulomb limit at the end of the integration yields
\begin{align}
\left(P_{\text{NR}} - P^{(0)}_{\text{NR}}\right)\big|_\text{Born} =  \frac{e^2 m^2_e T^2}{8 \pi^3} e^{-\beta 2m_e} \;,
\label{eq:p2born}
\end{align}
consistent with perturbative correction~\eqref{eq:p2NR}.  Thus, the Beth-Uhlenbeck formula~\eqref{eq:BU} reproduces the leading, ${\cal O}(e^2)$ result.

Next, we turn to a non-perturbative evaluation of~\eqref{eq:BU2}. Analytic expressions of the the binding energy and non-perturbative phase shifts do not exist for the screened Yukawa potential. We therefore approximate the Yukawa potential by the Hulthén potential,
\begin{align}
 V_H(r)= - \frac{ \alpha \kappa m_D e^{-\kappa m_D r}}{1-e^{-\kappa m_D r}} \;,\label{eq:VH}
\end{align}
where the choice of the parameter $\kappa=\pi^2/6$ is known to approximate the Yukawa potential particularly well in the vicinity of the first resonance (i.e., the appearance of the first $n=1, \ell=0$ zero-energy bound state)~\cite{Cassel:2009wt}. Even so, exact analytic expressions exist only for the $s$-wave%
\footnote{For higher partial waves, approximations of the centrifugal term are necessary to obtain analytic expressions. These exist, but are rather inaccurate because of quasi-bound states for $\ell >0$~\cite{Beneke:2024iev}.} 
(e.g.,~\cite{Tulin:2013teo, Mitridate:2017izz}),
\begin{align}
E_{n, \ell=0}&= - \frac{\alpha^2 m_e}{4 n^2} \left( 1- \kappa \epsilon_\phi n^2\right)^2 \;, \quad \text{for } m_D \leq \frac{\alpha m_e}{\kappa n^2} \;,\label{eq:HE}
\end{align}
and
\begin{align}
\delta_{\ell=0} &= \text{Arg} \frac{{\rm i} \Gamma(\lambda_+ + \lambda_- -2)}{\Gamma(\lambda_+)\Gamma(\lambda_-)} \;, \text{ where } \lambda_{\pm} =
\begin{cases}
	1+ {\rm i} \frac{\epsilon_v}{\kappa \epsilon_\phi} \pm \frac{\sqrt{\kappa \epsilon_\phi - \epsilon_v^2}}{\kappa \epsilon_\phi}, & \text{for } e^+ e^-\\
    1+ {\rm i} \frac{\epsilon_v}{\kappa \epsilon_\phi} \pm {\rm i} \frac{\sqrt{\kappa \epsilon_\phi + \epsilon_v^2}}{\kappa \epsilon_\phi},  &  \text{for } e^+ e^+, e^- e^-
\end{cases} \label{eq:Hphase}
\end{align}
with dimensionless quantities $\epsilon^2_v= E/(\alpha^2 m_e)$ and $\epsilon_\phi = m_D/(\alpha m_e)$.

\begin{figure}
    \centering
    \includegraphics[scale=0.573]
    {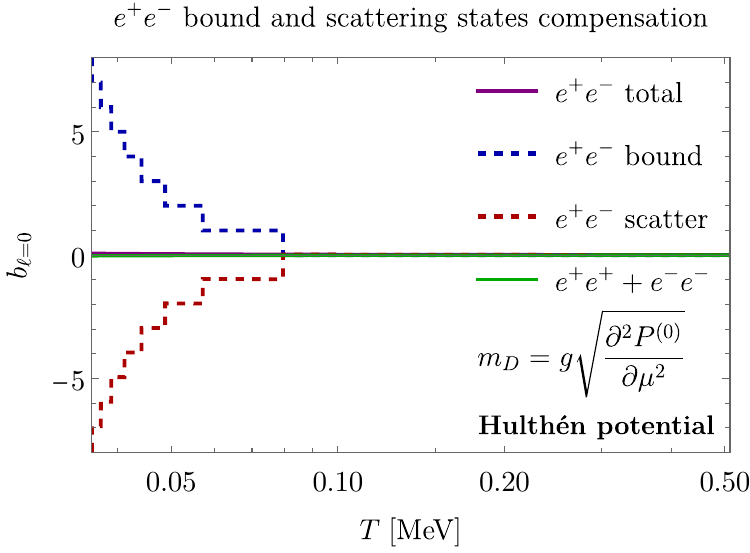}
     \includegraphics[scale=0.60]
    {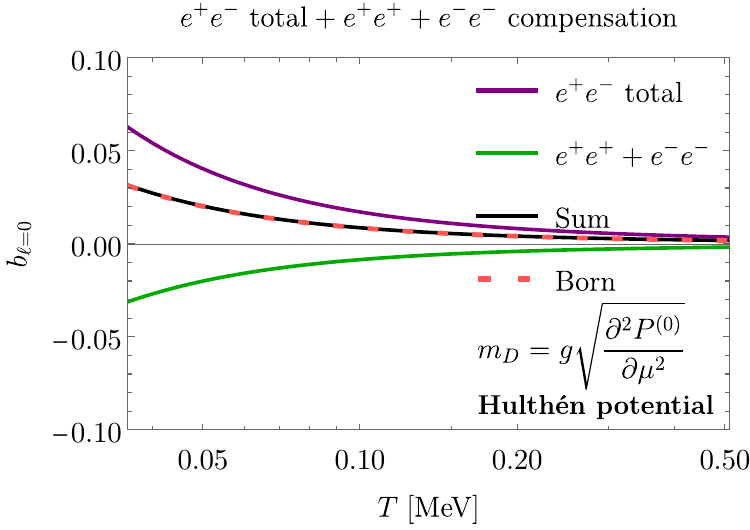}
    \caption{Breakdown of the various contributions to the coefficient $b_{\ell=0}$ in equation~\eqref{eq:BU2} as a function of temperature.}
    \label{fig:bl}
\end{figure}

Using equations~\eqref{eq:HE} and~\eqref{eq:Hphase}, we evaluate $b_{\ell=0}$ in \eqref{eq:BU2} semi-analytically for the Hulthén potential.
The results are shown in figure~\ref{fig:bl} for temperatures close to but less than the electron mass. In the left panel, the blue dashed line denotes the contribution from bound states, which exhibits a discrete jump every time a bound state is permitted to exist. Indeed, as the screening mass $m_D$ decreases exponentially with decreasing temperature, the number of permissible bound states also increases rapidly.  Similar jumps can be seen in the attractive scattering-state contributions---represented in figure~\ref{fig:bl} by the  magenta dashed line---but in the opposite direction.  Most interestingly, these discontinuities compensate precisely with those in the bound-state contribution, such that $b_{\ell=0}$ and hence the pressure and all other thermodynamic quantities in fact remain \emph{continuous}. The right panel zooms in on the sum of the $e^+e^-$-scattering and the bound-state contributions (purple solid line), which is indeed continuous and positive-valued.
In the high temperature limit (i.e., $\beta \rightarrow 0$), the $e^+ e^-$ bound- and scattering-state contributions cancel each other,
\begin{align}
\lim_{\beta \rightarrow 0} b_\ell = n_\ell + \frac{1}{\pi} \big[ \delta_\ell^{e^+ e^-}(E=\infty) - \delta_\ell^{e^+ e^-}(E=0) \big] =0 , 
\end{align}
where $n_\ell$ is the number of permissible bound states for a fixed $\ell$, and we have used Levinson's theorem.

Observe also the green solid line in the right panel of figure~\ref{fig:bl}.  This denotes the total contribution from the repulsive scattering states and runs negative, such that when summing up all contributions to $b_{\ell=0}$ in \eqref{eq:BU2}, there occurs another cancellation between the total attractive and total repulsive components, represented by the black solid line. This line is very close to the Born approximation computed using equation~\eqref{eq:phaseborn} for the Hulthén potential~\eqref{eq:VH} indicated by the red dashed line. We have checked that $b_{\ell=0}^\text{Born}$ for the Hulthén potential differs from the screened Yukawa case~\eqref{eq:bornyukawa} by at most 3\% in the temperature range shown in figure~\ref{fig:bl}. As the latter agrees with the ${\cal O}(e^2)$ perturbative computation of the pressure correction (see \eqref{eq:p2NR} versus~\eqref{eq:p2born}), we therefore conclude that perturbation theory is indeed under control if the potential is screened.


\subsection{Impact on \texorpdfstring{$N_\text{eff}$}{Neff}}
\label{sec:Neff}

Having now established that bound-state effects on the QED EoS is perturbatively small even at $T \ll m_e$, we are now in a position to estimate $\Delta N_\text{eff}$.  We adopt the usual approach of demanding comoving entropy conservation (e.g.,~\cite{Bennett:2019ewm}; see also equation~\eqref{eq:dNeff} and accompanying text), and estimate the change in the QED entropy,
\begin{align}
\delta s = \partial_T (P_\text{NR} - P_\text{NR}^{(0)}), \label{eq:dneffest}
\end{align}
from a non-ideal pressure correction $P_{\rm NR}- P_{\rm NR}^{(0)}$ using the non-perturbative Beth-Uhlenbeck formula~\eqref{eq:BU} for the Hulthén potential~\eqref{eq:VH} and only $s$-wave contributions.

\begin{figure}
    \centering
    \includegraphics[scale=0.61]
    {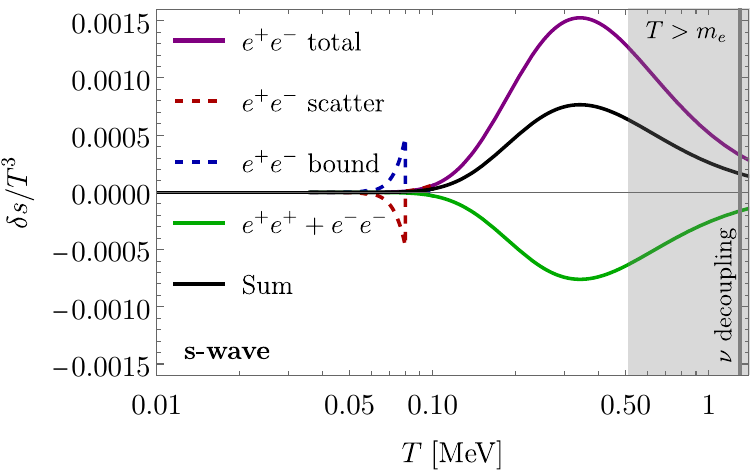}
     \includegraphics[scale=0.583]
    {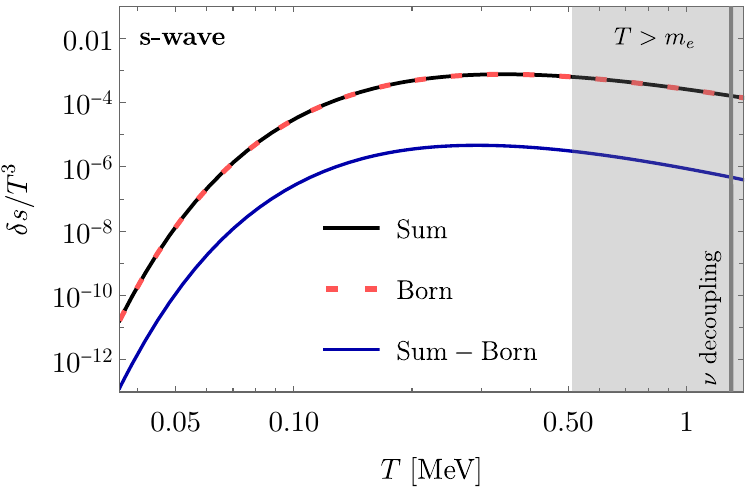}
    \caption{Change to the entropy, $\delta s$, as a function of temperature $T$.  {\it Left}: Breakdown of the contributions from various components.  {\it Right}: The sum of all contributions as derived from a full non-perturbative computation (black solid) and from the Born approximation (red dashed).  The difference between the non-perturbative and Born results (blue solid) remains substantially below the individual result, suggesting that perturbation theory is under control.}
    \label{fig:neffestimate}
\end{figure}

Figure~\ref{fig:neffestimate} show $\delta s/T^3$ as a function of $T$. The left panel displays a breakdown of the contributions from various components: attractive $e^+ e^-$ contributions from bound states (blue dashed) and scattering states  (magenta dashed), as well as the total repulsive contribution from $e^+ e^+$ and $e^- e^-$  (green solid). The bound-state contribution appears at $ T\simeq 80$~keV with a peak value of $\delta s/T^3 \sim 0.0005$, corresponding to the entropy of one Maxwellian-distributed species of $g_{\rm Ps}=4$ internal degrees of freedom.
Analogously to figure~\ref{fig:bl}, the contribution from attractive scattering states is almost as large in magnitude as their bound-state counterpart, but carries the opposite sign.  These two attractive contributions compensate each other almost exactly in summation (purple solid).  Furthermore, the total attractive contribution itself cancels to a good extent with the total repulsive contribution, leading to the sum total represented by the black solid line.  As in figure~\ref{fig:bl}, this total fully-non-perturbative $\delta s$ is remarkably close to what we would obtain from~\eqref{eq:BU} in the Born approximation (red dashed).  Indeed, as shown on the right panel of figure~\ref{fig:neffestimate}, subtracting the Born result from the total $\delta s$, the remainder---in blue, representing higher-order contributions---is two orders of magnitude smaller than the Born result.

To estimate $\Delta N_{\rm eff}$ due to bound states from these results, we note first of all that the ${\cal O}(e^2)$ correction to the QED EoS and its impact on $N^{\rm SM}_{\rm eff}$ have previously been considered in, e.g.,~\cite{Bennett:2019ewm}.  As we have established in sections~\ref{sec:eq} and~\ref{sec:BU}, in the non-relativistic limit, this correction coincides with the Born approximation of the Beth-Uhlenbeck formula.  Therefore, to avoid double counting, the true contribution of bound states to $N_{\rm eff}^{\rm SM}$ needs to be estimated from  the fully non-perturbative Beth-Uhlenbeck formula {\it minus} the Born piece---that is, the blue line in figure~\ref{fig:neffestimate}.

Secondly, following entropy conservation arguments, changes to $N_{\rm eff}^{\rm SM}$ in our scenario are ultimately linked to corrections to the QED entropy at neutrino decoupling, i.e.,
\begin{align}
\Delta N_{\rm eff} \simeq - 4 \left. \frac{\delta s}{s} \right|_{T=T_d} \simeq - \frac{180}{11 \pi^2} \left. \frac{\delta s}{T^3} \right|_{T=T_d}
\label{eq:dneffdsT3}
\end{align}
where we have approximated $s \simeq (11 \pi^2/45) T^3$ at the second equality, assuming massless electrons and positrons.  Given that $T_d \simeq 1.3~{\rm MeV} > m_e$, evaluation of~\eqref{eq:dneffdsT3} for bound state effects necessitates that we extrapolate to the relativistic regime outside of the validity region of the Beth-Uhlenbeck formula. Reading $\delta s/T^3$ off figure~\ref{fig:neffestimate} at $T=T_d$, we find $\Delta N_{\rm eff} \sim - 10^{-6}$---similar in magnitude to the relativistic ${\cal O}(e^4)$ correction identified in~\cite{Bennett:2019ewm} but with the opposite sign. We have also checked numerically that implementing the same EoS correction into the continuity equation~\eqref{eq:continuity} and solving it alongside a momentum-averaged Boltzmann equation (for the neutrinos) in the damping approximation yields a similarly small $\Delta N_{\rm eff} \sim - 6 \times 10^{-7}$.%
\footnote{This EoS correction can be straightforwardly implemented by making the following replacements in equation~\eqref{eq:continuity}: $G_1\to G_1 - (\tau^3/(2m_e^4))  \left[3(\bar{\rho}^{(4)} + \bar{P}^{(4)}+\tau \partial_\tau \bar{\rho}^{(4)}\right]$, and $G_2 \rightarrow G_2 - (\tau^4/(2 m_e^4)) \, \tau \partial_\tau \bar{\rho}^{(4)}$. Here, the rescaled energy density and pressure, $\bar{\rho}^{(4)}$ and $\bar{P}^{(4)}$, refer to the contribution that results after summing over all bound, scattering, and repulsive (i.e., $e^\pm e^\pm$) contributions and subtracting from it the already-known, perturbative $\mathcal{O}(e^2)$ correction.  For the neutrinos, the momentum-averaged Boltzmann equation in the damping approximation can be found in~\cite{Drewes:2024wbw}.}

Lastly, we note that for a perturbation theory that is under control, it is in principle possible to further subtract
from the blue line in figure~\ref{fig:neffestimate} the ${\cal O}(e^4)$ contribution, potentially yielding an even smaller $\delta s/T^3$ and hence $\Delta N_{\rm eff}$ due to bound states.   In this sense, our estimate of $\Delta N_{\rm eff} \sim - 10^{-6}$ above serves merely as an upper limit on the possible effect of bound states in the equilibrium scenario.  However, to perform such a subtraction would entail a consistent derivation of the non-relativistic limit of the relativistic QED pressure, as well as an expansion of the Beth-Uhlenbeck formula to higher orders, both of which are beyond the scope of the present work.  We also leave for future work a more complete and self-consistent determination of the full range of corrections due to bound states, including thermal width and the Salpeter correction, Bose-enhancement factors in the non-dilute case close to $T \lesssim m_e$, as well as summation over  partial waves, all of which are expected to be necessary in the temperature range of interest.


\section{Conclusions and outlook}
\label{sec:conclusions}

We have presented in this work a first assessment of how positronium states could affect the Standard Model prediction of the effective number of neutrinos, $\Neff^{\rm SM}$. We anticipate positronium to impact on $\Neff^{\rm SM}$ primarily through modifications to the QED EoS. 
Their impact via changes to the weak and electromagnetic particle scattering rates (e.g., enhanced annihilation cross section) is expected to be negligible.

A key challenge in quantifying the role of positronium in the QED EoS lies in the fact that the spectrum of bound states changes dynamically over the pertinent temperature regime because of in-medium effects such as screening and scatterings. This does not only shift the bound-state energies, but also dresses the states with thermal widths which make the transition to the continuum of scattering states gradual. To our knowledge, there is currently no known controlled approximation scheme that can reliably describe all potentially relevant effects across the entire temperature range of interest, covering the transition from the relativistic to the non-relativistic regime. In particular, we do not know to what degree the emergence of bound states implies a deviation from equilibrium in the early universe.

To systematically assess what different assumptions about equilibrium imply for $\Neff^{\rm SM}$, 
we adopt a simple model of the medium by modifying the Coulomb potential to a Yukawa potential via the introduction of a Debye mass $m_D$ for the photon.   In this approximation positronia can be treated as on-shell particles for which one can, e.g., unambiguously write down a set of Boltzmann equations to track their evolution in the early universe. We can then conceive of two possible positronium formation scenarios which differ mainly in their treatment of entropy evolution.

\paragraph{Out-of-equilibrium scenario} 
We consider first the viewpoint that positronium appears instantaneously in the spectrum of quantum states once the temperature drops to a certain value $T = T_{\rm Ps}^{\rm eq}$.
The blue line in figure~\ref{fig:tanh} shows the expected change in $\Neff^{\rm SM}$, $\Delta N_{\rm eff}$, under the assumption that the occupation number is then instantaneously driven towards equilibrium, as suggested by the large rate found in section~\ref{sec:poseqtemperature}.
The sudden transition from vanishing to equilibrium positronium occupation numbers at a single moment 
represents a sizeable deviation from thermal equilibrium and generates significant entropy.
Clearly, the corresponding $\Delta \Neff$ is exponentially dependent on $T^{\rm eq}_{\rm Ps}$ due to Boltzmann suppression of the positronium number density.
Identifying $T_{\rm Ps}^{\rm eq}$ with the highest temperature at which the real Yukawa potential can host a bound state, $T \simeq 75.6~\mbox{keV}$, would lead us to conclude that $|\Delta N_{\rm eff}| \lesssim 10^{-4}$, below the uncertainties of current precision $\Neff^{\rm SM}$ computations.

It is however possible that Debye screening is overcome in a more gradual manner, e.g., because of the finite thermal width of the quantum state.   We mimic the resulting effect on $\Delta \Neff$ by implementing a large transition parameter $\Delta T/T_{\rm Ps}^{\rm eq}$ symmetrically around $T_{\rm Ps}^{\rm eq}$ in our phenomenological modelling.  Because this effectively creates a small population of positronium at $T> T_{\rm Ps}^{\rm eq}$, a large transition parameter tends to drive up entropy production and hence $\Delta \Neff$ as indicated, for example, by the orange line in figure~\ref{fig:tanh}.

\paragraph{Equilibrium scenario} 
Here, we consider the case in which the QED sector {\it overall} is assumed to remain in thermal equilibrium at all times, even when crossing the existence threshold temperature of a positronium state.  In this limit, the primary contribution of positronium states to $\Neff^{\rm SM}$ should arise through non-perturbative corrections to the QED EoS around neutrino decoupling, $T_d \simeq 1$~MeV, i.e., a significantly higher temperature than the nominally highest temperature, $T \simeq 75.6~\mbox{keV}$, at which a bound state can exist.

To investigate this scenario, we first use the Beth-Uhlenbeck formula, a non-relativistic but non-perturbative expression for the pressure that accounts for photon-mediated interactions (``ladder diagrams'') to all orders in all two-particle bound and scattering states, 
to study thermodynamic quantities near the threshold temperature of the first $s$-wave positronium states. 
Figure~\ref{fig:neffestimate} shows that, while the bound-state and the attractive scattering-state contributions to the entropy each features a discontinuous jump at $T \simeq 75.6~\mbox{keV}$, their sum (as well as the total including repulsive contributions) varies smoothly across this threshold temperature.  This smoothness affirms that the appearance of bound states at low temperatures {\it need not} be accompanied by a sudden change in the overall QED entropy (although we emphasise that it {\it does not preclude} overall entropy production if bound state formation turns out to be disruptive, as in the out-of-equilibrium scenario).

Comparing the Born limit of the Beth-Uhlenbeck formula with a subpart of the leading-order non-ideal gas corrections computed using relativistic finite-temperature QED (the component~\eqref{eq:2a} previously called the ``logarithmic contribution'' in~\cite{Bennett:2019ewm}), we find agreement at $O(e^2)$ in the non-relativistic regime. This subpart is quadratic in the number density of electrons and positrons and has previously been shown to be negligible relative to the other leading-order contributions~\cite{Bennett:2019ewm}. However, the Beth-Uhlenbeck formula treats this quadratic number density term to all orders in the potential, which ultimately leads to the appearance of positronium states. 

A rather surprising result is that this non-perturbative treatment of potential insertions is in close agreement with the Born approximation of the Beth-Uhlenbeck formula in the relevant temperature regime, even at the formation threshold of the first $s$-wave positronium states (when summed over \emph{all} two-body states). In particular, the Born approximation captures the effects of scattering states well, as shown in the right panel of figure~\ref{fig:neffestimate}, indicating that perturbation theory can be trusted. 
This latter insight is important, as it supports the validity of the perturbative approach used in previous studies of non-ideal corrections to the QED EoS.  Subtracting the Born approximation from the Beth-Uhlenbeck formula and extrapolating  to the neutrino decoupling temperature, we find that in the equilibrium scenario the change in $N_{\rm eff}^{\rm SM}$ from the $s$-wave contribution
should be limited to $|\Delta N_{\rm eff}| \lesssim 10^{-6}$.  In view of the  proximity of this number in magnitude to the ${\cal O}(e^4)$ correction found in~\cite{Bennett:2019ewm}, it is probable that the result of~\cite{Bennett:2019ewm}  already subsumes the relevant effects.

\bigskip

Importantly, the discussion above---in both the out-of-equilibrium and the equilibrium scenario---ignores 
modifications to the bound-state properties due to scattering with the plasma constituents, such as Landau damping.  In the static limit, such effects can be captured by an imaginary component in the effective potential \eqref{VeffDef}.  Assuming that the sole effect of the imaginary potential is to push the bound-state formation threshold down to an even lower temperature, the out-of-equilibrium model for instantaneous production suggests $|\Delta \Neff| \simeq 3 \times 10^{-8}$, a number much smaller than the hard Debye cut-off estimate, $|\Delta \Neff| \simeq  10^{-4}$, discussed above.  In the equilibrium scenario, a self-consistent Beth-Uhlenbeck-like formulation that includes Landau damping (as well as the Salpeter correction) does not yet exist: we leave its exploration for future work.

In summary, a minute and transient population of positronium  in dynamical equilibrium with the rest of the QED plasma could in principle have existed
in roughly the same time frame during which the Standard-Model value of $\Neff$ is set. 
Uncertainties in relation to 
the dynamical evolution of bound-state properties 
and interaction rates in a QED plasma in the temperature range ${\cal O}(10) \ {\rm keV} \lesssim T \lesssim {\cal O}(100)$ keV prevent us from pinning down its exact contribution to $\Neff^{\rm SM}$.  
However, our current estimates suggest that the effect is likely to be $10^{-4}$ or smaller, i.e., not larger than other theoretical uncertainties, and an order of magnitude below the anticipated experimental sensitivity of CMB-S4.


\section*{Acknowledgements}

We are grateful to Georg Raffelt for triggering this research with an insightful question about the role of positronium.
We also thank Iason Baldes, Lorenzo de Ros, Mathias Garny, Jacopo Ghiglieri, and Georg Raffelt for useful discussions and comments on the manuscript. MaD acknowledges the hospitality of the Technical University of Munich in general and Bj\"orn Garbrecht in particular during part of the period when this research was performed. YG acknowledges the support of the French Community of Belgium through the FRIA grant No.~1.E.063.22F and thanks the UNSW School of Physics for its hospitality during part of this project. MK acknowledges financial support by DFG under grant KL 1266/11-1. GP and Y$^3$W are supported by the Australian Research
Council's Discovery Project (DP240103130) scheme. This work has been partially funded by the Deutsche Forschungsgemeinschaft (DFG, German Research Foundation) - SFB 1258 - 283604770.

\appendix

\section{Functions in the continuity equation}
\label{sec:continuitysupplement}

The continuity equation~\eqref{eq:continuity}, expressed in terms of the dimensionless variables $x,y,z$, contains the functions $J_{\pm}, Y_{\pm}, W_\pm$, and $G_{1,2}$.  We provide here their explicit forms~\cite{Mangano:2001iu}. At leading order:
\begin{subequations}
\begin{align}
    J_{\mathrm{\pm}}(\tau) =& \; \frac{1}{\pi^2} \int \d \omega \, \omega^2\frac{\exp\left(\sqrt{\tau^2+\omega^2}\right)}{\left[\exp\left(\sqrt{\tau^2+\omega^2}\right)\pm  1\right]^2},\\
    Y_{\mathrm{\pm}}(\tau) = & \; \frac{1}{\pi^2} \int \d \omega \, \omega^4\frac{\exp\left(\sqrt{\tau^2+\omega^2}\right)}{\left[\exp\left(\sqrt{\tau^2+\omega^2}\right)\pm  1\right]^2},\\
    W_\pm(\tau)= & \; \frac{1}{\pi^2}\int_0^{\infty} {\rm d}\omega\, \omega^2\frac{\sqrt{\omega^2+\tau^2}}{\exp(\sqrt{\omega^2+\tau^2})\pm 1},\\
  K_\pm(\tau)= & \; \frac{1}{\pi^2}\int_0^{\infty} {\rm d}\omega\, \frac{\omega^2}{\sqrt{\tau^2 + \omega^2}} \frac{1}{\exp(\sqrt{\omega^2+\tau^2})\pm 1},
 \end{align}
 \end{subequations}
where $' \equiv \partial/\partial \tau$, and the $+/-$ signs pertain to the Fermi-Dirac and the Bose-Einstein distributions respectively.
At ${\cal O}(e^2)$, the functions $G_{1,2}(\tau)$ contain two components each, i.e., $G_i = G_{i,\slashed{\ln}} + G_{i,\ln}$, given by~\cite{Mangano:2001iu,Bennett:2019ewm}
\begin{subequations}
 \begin{align} 
G_{1, \slashed{\ln}}(\tau) = & \; 2 \pi \alpha \Bigg[\frac{1}{\tau}
\left(\frac{K(\tau)}{3} + 2 K^2(\tau)- \frac{J(\tau)}{6}- K(\tau)J(\tau)\right) \\
\nonumber
&\; + \left(\frac{K'(\tau)}{6}- K(\tau) K'(\tau) + \frac{J'(\tau)}{6} + J'(\tau) K(\tau) + J(\tau) K'(\tau)\right) \Bigg] , \\
G_{1, \ln}(\tau) = &\; \frac{\alpha}{4 \pi^3} \tau \iint_0^\infty {\rm d} \omega \; {\rm d} \tilde{\omega} \frac{\omega}{\sqrt{\omega^2+\tau^2}} \frac{\tilde{\omega}}{\sqrt{\tilde{\omega}^2+\tau^2}}  \ln \left|\frac{\omega+ \tilde{\omega}}{\omega - \tilde{\omega}} \right| \nonumber\\
&\; \times  \Bigg\{ - \frac{1}{2} \left(\frac{\omega^2}{\omega^2+\tau^2} + \frac{\tilde{\omega}^2}{\tilde{\omega}^2+\tau^2} \right) n_D \tilde{n}_D + \left[\frac{\tilde{\omega}^2+6 \tau^2}{\tau} + \frac{\tau (\tilde{\omega}^2-\omega^2)}{\omega^2 + \tau^2}\right] n_D\tilde{n}_D' \\
&\; \qquad
+(\tilde{\omega}^2+\tau^2) \left( n_D'\tilde{n}_D'+ n_D \tilde{n}_D'' \right)
\Bigg\}, \nonumber \\
G_{2, \slashed{\ln}}(\tau) = & \; - 8 \pi \alpha 
\left(\frac{K(\tau)}{6} +\frac{J(\tau)}{6} - \frac{1}{2} K^2(\tau)+K(\tau)J(\tau)\right) \\
\nonumber
&\; + 2 \pi \alpha\Bigg[\frac{K'(\tau)}{6}- K(\tau) K'(\tau) + \frac{J'(\tau)}{6} + J'(\tau) K(\tau) + J(\tau) K'(\tau) \Bigg] , \\
G_{2,\ln}(\tau) = &\; \frac{\alpha}{4 \pi^3} \tau^2 \iint_0^\infty {\rm d} \omega \; {\rm d} \tilde{\omega} \frac{\omega}{\sqrt{\omega^2+\tau^2}} \frac{\tilde{\omega}}{\sqrt{\tilde{\omega}^2+\tau^2}}  \ln \left|\frac{\omega+ \tilde{\omega}}{\omega - \tilde{\omega}} \right| \\
&\; \times \Bigg\{\frac{(\omega^2 + \tau ^2)(\tilde{\omega}^2+\tau^2)}{\tau^2}
n_D' \tilde{n}_D' 
+ n_D (\tilde{\omega}^2 + \tau^2) \left[ \frac{2}{\tau} \tilde{n}_D' + \frac{\tilde{\omega}^2+\tau^2}{\tau^2} \tilde{n}_D''\right]
\Bigg\}, \nonumber
\end{align}
\end{subequations}
where $n_D (\omega, \tau)=  [\exp(\sqrt{\omega^2+\tau^2}) + 1]^{-1}$ is the Fermi-Dirac distribution, $\tilde{n}_D \equiv n_D(\tilde{\omega},\tau)$, and $'' \equiv \partial^2/\partial \tau^2$.
Expressions for $G_{1,2}(\tau)$ at ${\cal O}(e^3)$ can be found in reference~\cite{Bennett:2019ewm}.


\section{Beth-Uhlenbeck formula for a non-relativistic \texorpdfstring{$e^+ e^-$}{e+e-} plasma}
\label{app:bu}

We summarise here a derivation of the Beth-Uhlenbeck formula~\eqref{eq:BU}; see also \cite{kremp2005quantum, bookEB}.


\subsection{Charging formula}

Given the non-relativistic Hamiltonian~\eqref{eq:ham}, all thermodynamics quantities relevant for $\Delta N_{\rm eff}$ can be established from the associated partition function $Z$ and hence pressure $P= T \ln Z$.  To compute the partition function, we make use of the charging formula, 
\begin{align}
P{\cal V}-P^{(0)}{\cal V} &= T [\ln Z - \ln Z_0] = T \int_0^1 \text{d} \lambda\; \partial_\lambda \ln Z_{\lambda}= - \int_0^1 \text{d} \lambda \; \frac{\text{Tr}[e^{-\beta (H_0+\lambda H_\text{int})} H_\text{int}]}{Z_\lambda} \;,
\label{eq:pv}
\end{align}
which is a formal trick to evaluate the pressure in a non-perturbative way. Here, $Z_\lambda=\text{Tr}[e^{-\beta (H_0+\lambda H_\text{int} )}]$, and we have omitted the sub- and superscripts ``$\text{NR}$'' for notational simplicity. In the language of the real-time formalism---also called the closed-time path (CTP) formalism---one can identify from the structure of $H_\text{int}$ in equation~\eqref{eq:ham} three equal-time equilibrium correlators,
\begin{subequations}
\label{eq:g}
\begin{align}
G^{++--}_{\eta \xi,\lambda}(t\mathbf{x},\mathbf{y};t\mathbf{x},\mathbf{y})&\equiv \langle \eta^{\dagger}_\sigma(\mathbf{x}) \xi_{\sigma^\prime}(\mathbf{y})  \xi^{\dagger}_{\sigma^\prime}(\mathbf{y})\eta_{\sigma}(\mathbf{x}) \rangle_\lambda \;, \label{eq:spinetaxi}\\
G^{++--}_{\eta \eta,\lambda}(t\mathbf{x},\mathbf{y};t\mathbf{x},\mathbf{y}) &\equiv \langle \eta^{\dagger}_\sigma(\mathbf{x})  \eta^{\dagger}_{\sigma^\prime}(\mathbf{y})   \eta_{\sigma^\prime}(\mathbf{y}) \eta_{\sigma}(\mathbf{x}) \rangle_\lambda \;,\label{eq:spinetaeta}\\
G^{++--}_{\xi \xi,\lambda}(t\mathbf{x},\mathbf{y};t\mathbf{x},\mathbf{y}) &\equiv \langle \xi_{\sigma}(\mathbf{x}) \xi_{\sigma^\prime}(\mathbf{y})  \xi^{\dagger}_{\sigma^\prime}(\mathbf{y}) \xi^{\dagger}_\sigma(\mathbf{x})\rangle_\lambda \;, \label{eq:spinetaxi3}
\end{align}
\end{subequations}
where $\langle ... \rangle_\lambda = \text{Tr}[e^{-\beta (H_0+\lambda H_\text{int})}...]/Z_\lambda $, all field operators on the right are evaluated at the same time $t$, and the superscript ``$++--$'' denotes a certain operator ordering (of a more general correlator on the CTP contour; the interested reader may consult~\cite{Binder:2018znk} for the convention, although it is of no relevance here).

In terms of the correlators~\eqref{eq:g}, the pressure~\eqref{eq:pv} can now be written as
\begin{align}
P {\cal V}-P^{(0)} {\cal V } = \int_0^1 \text{d} \lambda\;  \frac{1}{2} \sum_{\sigma,\sigma^\prime} \int \text{d}^3\mathbf{x} \, \text{d}^3
\mathbf{y} \; V(|\mathbf{x}-\mathbf{y}|) \bigg[ G^{++--}_{\eta \eta,\lambda}(t\mathbf{x},\mathbf{y};t\mathbf{x},\mathbf{y})+ G^{++--}_{\xi \xi,\lambda} -2 G^{++--}_{\eta \xi,\lambda} \bigg ], 
\label{eq:chargingcorrelators}
\end{align}
where the spin indices of the correlators are implicit and set by equations~\eqref{eq:g}.   We can further connect these correlators to spectral functions via the Kubo-Martin-Schwinger (KMS) relation. Taking the dilute limit and expressed in Wigner coordinates and momentum space, these relations are (see, e.g.,~\cite{Binder:2018znk}, chapter~V)
\begin{align}
	G^{++--}_{\bullet \bullet,\lambda} (t,\mathbf{x},\mathbf{y};t,\mathbf{x},\mathbf{y}) = 
	e^{- \beta 2 m_e}
	\int \frac{\text{d}^3 \mathbf{K}}{(2 \pi)^3} e^{- \beta \mathbf{K}^2 / (4 m_e)}
	\int_{-\infty}^{\infty} \frac{\text{d}E}{(2\pi)} e^{- \beta E} G^\rho_{\bullet \bullet,\lambda} (E; \mathbf{r},\mathbf{r}),
	\label{eq:coll_th}
\end{align}
where $\bullet = \eta,\xi$, $K$ is the centre-of-mass momentum, and $\mathbf{r}$ is the relative distance. Observe that $E$ runs from $-\infty$ to $+\infty$: 
a positive value of $E=m_e v^2/4$ denotes the relative kinetic energy with relative velocity $v$, while a negative value corresponds to the binding energy once the spectral function $G^\rho$ has support. 

Then, recasting~\eqref{eq:chargingcorrelators} in terms of~\eqref{eq:coll_th} and performing the integration over ${\bf K}$ and the centre-of-mass position (the latter integration yields the volume~${\cal V}$), 
the charging formula~\eqref{eq:pv} now reads 
\begin{align}
P-P^{(0)} &= \left( \frac{2m_e T}{2 \pi} \right)^{3/2} e^{- \beta 2 m_e} \frac{1}{2}  \sum_{\sigma,\sigma^\prime} \int_0^1 \text{d} \lambda  \int \text{d}^3 \mathbf{r} V(r)  \times  \label{eq:pinterm}\\
&\int_{-\infty}^{\infty} \frac{\text{d}E}{(2\pi)} e^{- \beta E} \bigg[ G^\rho_{\eta \eta,\lambda} (E; \mathbf{r},\mathbf{r}) + G^\rho_{\xi\xi,\lambda} (E; \mathbf{r},\mathbf{r}) - 2 G^\rho_{\eta \xi,\lambda} (E; \mathbf{r},\mathbf{r}) \bigg] \;. \nonumber
\end{align}
We have not yet specified the form of the spectral functions~$G^\rho$; these can be obtained from the imaginary part of the retarded correlator, to be derived next.


\subsection{Non-relativistic four-point correlators}

Our starting point to estimate the retarded four-point correlators is non-relativistic QED~\cite{Caswell:1985ui}, where the non-relativistic effective action  $S= S^0 + S^\text{int}$ in vacuum is composed of
\begin{align}
	S^0 &= \sum_\sigma \int \text{d}^4 x \; 
		  \eta^\dag_\sigma(x) \left[ {\rm i} \partial_{t} + \frac{\nabla^2}{2 m_e}  \right] \eta_\sigma(x) +
		\xi^\dag_\sigma(x) \left[{\rm i} \partial_{t} - \frac{\nabla^2}{2 m_e}  \right] \xi_\sigma(x) \;,  \nonumber\\
   S^\text{int}  & = \frac{1}{2} \sum_{\sigma,\sigma^\prime} \int \text{d}^4x \, \text{d}^3
\mathbf{y} \; V(|\mathbf{x}-\mathbf{y}|)  \label{eq:nr_eft_u1} \\ 
& \; \times \bigg[ \eta^{\dagger}_\sigma(\mathbf{x}) \eta_{\sigma}(\mathbf{x})  \eta^{\dagger}_{\sigma^\prime}(\mathbf{y}) \eta_{\sigma^\prime}(\mathbf{y}) +  \xi^{\dagger}_\sigma(\mathbf{x}) \xi_{\sigma}(\mathbf{x})  \xi^{\dagger}_{\sigma^\prime}(\mathbf{y}) \xi_{\sigma^\prime}(\mathbf{y}) +2 \eta^{\dagger}_\sigma(\mathbf{x}) \eta_{\sigma}(\mathbf{x})  \xi^{\dagger}_{\sigma^\prime}(\mathbf{y}) \xi_{\sigma^\prime}(\mathbf{y}) \bigg] \;, \nonumber
\end{align}
and we focus on potential exchanges only. Note the different operator ordering compared to equation~\eqref{eq:ham}. 

We wish to derive from the action~\eqref{eq:nr_eft_u1} the equations of motion for the four-point correlators in vacuum,
\begin{equation}
\begin{aligned}
G_{\eta \eta}(x,y;z,w) &=\langle T \eta_{\sigma_1}(x) \eta_{\sigma_2}(y) \eta^\dagger_{\sigma_3}(w) \eta^\dagger_{\sigma_4}(z) \rangle \;, \\
G_{\eta \xi}(x,y;z,w) &=\langle T \eta_{\sigma_1}(x) \xi_{\sigma_2}^\dagger (y) \xi_{\sigma_3}(w) \eta^\dagger_{\sigma_4}(z) \rangle \;.
\end{aligned}
\end{equation}
In integral form, these equations of motion are given by
\begin{subequations}
\begin{align}
G_{\eta \eta}(t \mathbf{x}\mathbf{y}, t^\prime \mathbf{z} \mathbf{w})&=  G_\eta(t\mathbf{x},t^\prime \mathbf{z}) G_\eta(t\mathbf{y},t^\prime \mathbf{w}) -  G_\eta(t\mathbf{x},t^\prime \mathbf{w}) G_\eta(t\mathbf{y},t^\prime \mathbf{z}) \label{eq:eom1} \\
&+{\rm i} \int \text{d}\bar{t} \, \text{d}^3 \mathbf{a}\,  \text{d}^3 \mathbf{b}\;  G_\eta(t \mathbf{x},\bar{t} \mathbf{a})G_\eta(t \mathbf y,\bar{t} \mathbf{b}) V(\mathbf{a}-\mathbf{b})G_{\eta\eta}(\bar{t}\mathbf{a}\mathbf{b},t^\prime \mathbf{z} \mathbf{w}) \,, \nonumber \\
G_{\eta \xi}(t \mathbf{x}\mathbf{y}, t^\prime \mathbf{z} \mathbf{w})&=  G_\eta(t\mathbf{x},t^\prime \mathbf{z}) G_\xi(t\mathbf{y},t^\prime \mathbf{w}) \label{eq:eom2} \\
&-{\rm i} \int \text{d}\bar{t} \, \text{d}^3 \mathbf{a} \, \text{d}^3 \mathbf{b} \; G_\eta(t \mathbf{x},\bar{t} \mathbf{a})G_\xi(t \mathbf y,\bar{t} \mathbf{b}) V(\mathbf{a}-\mathbf{b})G_{\eta\eta}(\bar{t}\mathbf{a}\mathbf{b},t^\prime \mathbf{z} \mathbf{w}) \;, \nonumber
\end{align}
\end{subequations}
where we have truncated the correlator hierarchy at the four-point function level. In differential form and approximating the two-point functions by their free counterparts, they appear in Wigner coordinates as
\begin{subequations}
\begin{align}
\left[    E+{\rm i}\epsilon + \frac{\nabla^2_{\mathbf{r}}}{m_e}+ V(\mathbf{r}) \right]G_{\eta \eta}(E;\mathbf{r},\mathbf{r}^{\prime}) &= {\rm i} \delta_{\sigma_1 \sigma_4}\delta_{\sigma_2 \sigma_3}\delta^{(3)}(\mathbf{r} - \mathbf{r}^{\prime})-{\rm i} \delta_{\sigma_1 \sigma_3}\delta_{\sigma_2 \sigma_4}\delta^{(3)}(\mathbf{r} + \mathbf{r}^{\prime}), \label{eq:appEOM1} \\
\left[    E+ {\rm i}\epsilon + \frac{\nabla^2_{\mathbf{r}}}{m_e} - V(\mathbf{r}) \right]G_{\eta \xi}(E;\mathbf{r},\mathbf{r}^{\prime}) &= {\rm i} \delta_{\sigma_1 \sigma_4}\delta_{\sigma_2 \sigma_3}\delta^{(3)}(\mathbf{r} - \mathbf{r}^{\prime}) \;,\label{eq:appEOM2} 
\end{align}
\end{subequations}
and admit Green's function solutions 
\begin{subequations}
\begin{align}
 G_{\eta \eta}(E;\mathbf{r},\mathbf{r}^{\prime}) &= {\rm i}  \int \frac{\text{d}^3\mathbf{p}}{(2\pi)^3}\frac{\delta_{\sigma_1 \sigma_4}\delta_{\sigma_2 \sigma_3}\psi_\mathbf{p}(\mathbf{r}) \psi^\star_\mathbf{p}(\mathbf{r}^\prime) -\delta_{\sigma_1 \sigma_3}\delta_{\sigma_2 \sigma_4}\psi_\mathbf{p}(\mathbf{r}) \psi^\star_\mathbf{p}(-\mathbf{r}^\prime)}{E-p^2/m_e + {\rm i} \epsilon} \;, \label{eq:soletaeta}\\
 G_{\eta \xi}(E;\mathbf{r},\mathbf{r}^{\prime}) &={\rm i} \delta_{\sigma_1 \sigma_4}\delta_{\sigma_2 \sigma_3}   \sumint_\gamma \frac{\psi_\gamma(\mathbf{r}) \psi^\star_\gamma(\mathbf{r}^\prime)}{E-E_\gamma + {\rm i} \epsilon} \;,\label{eq:soletaxi}
\end{align}
\end{subequations}
expressed here in spectral representation.  Observe that the wavefunctions $\psi$ are the solution of the homogenous (Schr\"odinger) equation and fulfil the completeness relation and normalisation
\begin{subequations}
\begin{align}
\sumint_\gamma \psi_\gamma(\mathbf{r}) \psi^\star_\gamma(\mathbf{r}^\prime) &= \delta^{(3)}(\mathbf{r}-\mathbf{r}^\prime) \;, \label{eq:comp} \\ 
\int \text{d}^3 \mathbf{r} \; \psi_\mathbf{p}(\mathbf{r}) \psi^\star_{\mathbf{p}^\prime}(\mathbf{r})&=(2\pi)^3 \delta^{(3)}(\mathbf{p}-\mathbf{p}^\prime)\;, \\
\int \text{d}^3 \mathbf{r} \; \psi_{n \ell m}(\mathbf{r}) \psi^\star_{n^\prime \ell^\prime m^\prime}(\mathbf{r}) &= \delta_{n n^\prime} \delta_{\ell \ell^\prime} \delta_{m m^\prime} \;.
\end{align}
\end{subequations}
Starting from equation~\eqref{eq:comp},  it is straightforward to verify that equations~\eqref{eq:soletaeta} and \eqref{eq:soletaxi} are the solutions to \eqref{eq:appEOM1} and \eqref{eq:appEOM2}, respectively.


\subsection{Integral reductions}

To evaluate the charging formula~\eqref{eq:pinterm}, we use the solutions~\eqref{eq:soletaeta} and~\eqref{eq:soletaxi}
to reduce the number of integrals. We perform first a partial integration,
\begin{align}
&\int \! \text{d} E e^{-\beta E }\! \int_0^1  \text{d} \lambda \! \int \text{d}^3 \mathbf{r}   V(\mathbf{r})  G^{\rho}_{\bullet \bullet,\lambda}(E;\mathbf{r},\mathbf{r})=
T \!\! \int \! \text{d} E e^{-\beta E} \! \int_0^1  \text{d} \lambda \!  \int \text{d}^3 \mathbf{r} \; V(\mathbf{r})  \partial_E  G^{\rho}_{\bullet \bullet,\lambda}(E;\mathbf{r},\mathbf{r}) . 
\label{eq:partialint}
\end{align}
Here, the spectral function $G^\rho$ can be computed from $G^\rho_{\bullet \bullet, \lambda} = G^R_{\bullet \bullet, \lambda} - G^A_{\bullet \bullet, \lambda}$, where $G^R$ and $G^A$  are the retarded and the advanced correlators, respectively.  In vacuum, the retarded correlator $G^R_{\bullet \bullet}$ is equivalent to the time-ordered correlators~\eqref{eq:soletaeta} or~\eqref{eq:soletaxi} for $\bullet \bullet = \eta \eta$ and $\eta \xi$, respectively, while its advanced counterpart has $-{\rm i}\epsilon$ instead of $+ {\rm i} \epsilon$ in the denominator. 

Noting that the energy derivative acting on the retarded correlator can be written as
\begin{align}
\partial_E \bra{ \mathbf{r} } \frac{{\rm i} V}{E-h_0 \mp \lambda V + {\rm i} \epsilon } \ket{ \mathbf{r} } = \mp \partial_\lambda \bra{ \mathbf{r} } \frac{{\rm i}}{E-h_0 \mp \lambda V + {\rm i} \epsilon } \ket{ \mathbf{r} } \;,
\end{align}
and analogously for the advanced correlator, the $\lambda$-integration in~\eqref{eq:partialint} evaluates trivially to
\begin{align}
\int_0^1 \text{d} \lambda  \int \text{d}^3 \mathbf{r} V(r) \partial_E G^{\rho}_{\bullet \bullet, \lambda}(E;\mathbf{r},\mathbf{r}) = \mp \int \text{d}^3 \mathbf{r} \bigg[ G^{\rho}_{\bullet \bullet}(E;\mathbf{r},\mathbf{r}) -G^{\rho}_{\bullet \bullet,0}(E;\mathbf{r},\mathbf{r})  \bigg] \;,
\label{eq:lambdaint}
\end{align}
where $G^{\rho}_{\bullet \bullet,0}$ and $G^{\rho}_{\bullet \bullet}$ are the free and the fully interacting ($\lambda=1$) spectral function, respectively, and the ``-'' (``+'') sign applies to the attractive (repulsive) case.  
 Then, we arrive at
\begin{align}
\beta P-\beta P^{(0)} =& \left( \frac{2m_e T}{2 \pi} \right)^{3/2} e^{- \beta 2 m_e}   \int_{-\infty}^\infty \frac{\text{d}E}{(2\pi)} e^{- \beta E} \times \label{eq:interPP}\\ 
& \sum_{\sigma,\sigma^\prime} \int \text{d}^3 \mathbf{r} \bigg[ G^\rho_{\eta \xi} (E; \mathbf{r},\mathbf{r}) - G^\rho_{\eta \xi,0} (E; \mathbf{r},\mathbf{r})  + \frac{1}{2} \left( G^\rho_{\eta \eta} -G^\rho_{\eta \eta,0} + G^\rho_{\xi\xi} - G^\rho_{\xi\xi,0} \right) \bigg] \; \nonumber
\end{align}
as an intermediate result.

To perform the radial integration, consider first the $E<0$ case.  Here, only the interacting attractive case is non-vanishing, giving the positronium bound states. The integrand on the right-hand side of~\eqref{eq:lambdaint} is therefore
\begin{align}
 \bigg[G^{\rho}_{\eta \xi}(E;\mathbf{r},\mathbf{r}) -G^{\rho}_{\eta \xi,0}(E;\mathbf{r},\mathbf{r}) \bigg] \bigg|_{E<0} = G^{\rho}_{\eta \xi}(E;\mathbf{r},\mathbf{r}) \big|_{E<0}= \delta_{\sigma \sigma} \delta_{\sigma^\prime \sigma^\prime}\sum_{\mathcal{B}} | \psi_{\mathcal{B}}(\mathbf{r}) |^2 (2\pi) \delta(E- E_\mathcal{B}) \;. \label{eq:boundG}
\end{align}
 Then, inserting \eqref{eq:boundG} into equation~\eqref{eq:interPP} and performing the radial and energy integration (noting that all bound-state wavefunctions are normalised to unity under radial integration) and spin summation yield the bound-state part of the Beth-Uhlenbeck formula~\eqref{eq:BU}.

For $E>0$ scattering states, decomposing the Green's functions into partial waves,
\begin{align}
 G_{\bullet \bullet}(E; \mathbf{r}, \mathbf{r}^\prime) &= \sum_\ell \frac{2\ell+1}{4 \pi} P_\ell(\cos \theta) G_{\bullet \bullet,\ell}(E, r,r^\prime) \,,
 \end{align}
leaves us with the radial integral
\begin{align}
\int_0^\infty \text{d}r \big[ u_{k \ell}(r) u_{k \ell}(r) - u_{k \ell}^0(r) u_{k \ell}^0(r) \big] = \frac{1}{\pi} \frac{\text{d}E}{ \text{d} k } \frac{\text{d}}{\text{d}E} \delta_\ell(E) \;,
\label{eq:radiale>0}
\end{align}
where we have used the Schr\"odinger equation for the radially reduced wavefunction $u_{k \ell}$ and its asymptotic behaviour at large $r$; see chapter 3.5 in~\cite{bookEB} for details. Equation~\eqref{eq:radiale>0} is the origin of the phase shifts $\delta_\ell$ in the Beth-Uhlenbeck formula~\eqref{eq:BU}. The non-trivial factor $(1-(-1)^\ell/2)/2$ preceding the phases of the repulsive degenerate states follows on the other hand from the spin structure in~\eqref{eq:soletaeta} and the partial wave decomposition of $\psi_\mathbf{p}(\mathbf{r}) \psi^\star_\mathbf{p}(-\mathbf{r}^\prime)$, which yields $P_\ell(-\cos \theta)=(-1)^\ell P_\ell(\cos \theta)$.


\section{Non-relativistic Feynman rules in momentum space}
\label{app:FR}

The two-point function of non-relativistic electrons on the CTP contour $\mathcal{C}$ is defined as
\begin{align}
G_{\eta}(x,y) \equiv \langle T_\mathcal{C} \eta_\sigma(x) \eta^\dagger_{\sigma^\prime}(y) \rangle.
\end{align}
In Wigner coordinates, the corresponding Fourier-transformed correlator is given by
\begin{align}
G_{\eta}(x-y) &= \int \frac{\text{d}^4p}{(2\pi)^4} e^{-{\rm i}p(x-y)} G_{\eta}(p) \;.
\end{align}
Of special interest is the free ``$+-$'' component, which in equilibrium can be expressed via the KMS relation as
\begin{equation}
\begin{aligned}
 G_{\eta,0}^{+-}(p) &=  G_{\eta,0}^{\rho}(p) e^{-\beta(m_e + p^0)} \;,\\
 G_{\eta,0}^{\rho}(p) &= \delta_{\sigma \sigma^\prime}2 \pi \delta(p^0-\mathbf{p}^2/(2m_e))\;,
\end{aligned}
\end{equation}
where $G^\rho_{\eta,0}$ is the free single electron spectral function. These expressions are used in equation~\eqref{eq:p2NR}.

\bibliographystyle{JHEP}
\bibliography{main}

\end{document}